\documentclass[12pt]{iopart}

\usepackage{graphicx}

\begin{document}

\title{Energy Level Statistics of Quantum Dots}
\author{Chien-Yu Tsau, Diu Nghiem, and Robert Joynt}
\address{University of Wisconsin-Madison, Madison, WI 53706, USA}
\author{J.Woods Halley}
\address{School of Physics and Astronomy, University of Minnesota,\\
 Minneapolis, MN 55455, USA }
\begin{abstract}
We investigate the charging energy level statistics of disordered
interacting electrons in quantum dots by numerical calculations
using the Hartree approximation. The aim is to obtain a global 
picture of the statistics as a function of disorder and interaction
 strengths. We find Poisson statistics at very strong disorder,
 Wigner-Dyson statistics for weak disorder and interactions, and a
 Gaussian intermediate regime. These regimes are as expected from
 previous studies and fundamental considerations, but we also find
 interesting and rather broad crossover regimes. In particular,
 intermediate between the Gaussian and Poisson regimes we find a
 two-sided exponential distribution for the energy level spacings.
 In comparing with experiment, we find that this distribution may
be realized in some quantum dots.\\
\end{abstract}

\section{Introduction} \label{sec:introduction}

Understanding energy level statistics (ELS) of quantum many-body systems is
a fundamental and intriguing challenge. Wigner first proposed the
statistical method in order to understand the excitation energies of nuclei,
and developed the mathematics of random matrix theory (RMT) to do the
calculations \cite{wigner}. This idea has been very successful in
elucidating experimental data in nuclear spectroscopy \cite{bohigas}. The
advent of artificially constructed finite interacting quantum systems
provides an opportunity to test these ideas again \cite{gorkov}. Quantum
dots are the system of choice today, and indeed RMT is useful in describing
transport and excitation energies in dots \cite{alhassid}. Dots have the
additional feature that the particle number can be changed in a controlled
fashion, and one can investigate a somewhat different quantity, the change
in ground state energy when a particle is added to the dot. This
distribution of level spacings when the particle number is changed will be
termed the charging energy level statistics (CELS). Surprisingly, these
statistics of this quantity do not follow RMT at all \cite{sivan}.

The CELS\ is measured as follows. In the Coulomb blockade regime, the
conductance of a dot is highly resonant, with a sharp peak when the chemical
potential difference of the leads is equal to the difference $%
E_{G}(N+1)-E_{G}(N)$, where $E_{G}(N)$ is the total ground state energy of
the dot with $N$ particles. Since the particle number can be varied by
adjusting the gate voltage, the quantity $\Delta _{2}(N)\equiv
E_{G}(N+1)-2E_{G}(N)+E_{G}(N-1)$ can be measured by recording the spacing of
adjacent conductance peaks on the graph of conductance vs. gate voltage. $%
\Delta _{2}$ fluctuates as the particle number is varied. By measuring it
for many different dot fillings, a probability distribution $P(\Delta _{2})$
can be built up, and it is this distribution that is compared to theory.

The simplest way to apply RMT\ to the CELS is via the constant interaction
model, which goes as follows \cite{jalabert}. Let the dot have charge $Q=-Ne$
and capacitance $C$. Now assume that one may separate the energy into a
non-fluctuating (\textquotedblleft constant\textquotedblright ) energy of
interaction $E_{C}(N)=Q^{2}/2C$ and a fluctuating part $E_{f}(N)$. Then 
\begin{eqnarray*}
\Delta _{2}(N) &=&(e^{2}/2C)[(N+1)^{2}-2N^{2}+(N-1)^{2}] \\
& &+E_{f}(N+1)-2E_{f}(N)+E_{f}(N-1) \\
&=& e^{2}/C+E_{f}(N+1)-2E_{f}(N)+E_{f}(N-1).
\end{eqnarray*}
Further assume that RMT can be applied to $E_{f}(N)$ and the conclusion is
that $\Delta _{2}(N)-e^{2}/C$ should follow Wigner-Dyson statistics. That
is, if $\Delta _{2}(N)$ is measured for many $N$ and a histogram is built
up, the shape of the histogram, when normalized to unit area, should
converge (to a very good approximation) to the form%
\begin{equation}  \label{eq:wdstat}
P_{WD}(\Delta _{2}-e^{2}/C) = \left\{ 
\begin{array}{l}
0, \,\,\,\,\textrm{ if } \Delta _{2}-e^{2}/C<0 \\ 
\frac{\pi }{2s^{2}}\left( \Delta _{2}-e^{2}/C\right) \exp \left[ \frac{-\pi 
}{4}(\frac{\Delta _{2}-e^{2}/C}{s})^{2}\right], \, \textrm{otherwise.}%
\end{array}
\right.
\end{equation}
This is sometimes called the CI (constant interaction) + RMT model. The
prediction of Eq. (\ref{eq:wdstat}) is in stark contradiction to experiments
on the CELS, which show a $P(x)$ that is usually approximately Gaussian
instead of having the asymmetric shape predicted by Eq. (\ref{eq:wdstat}) 
\cite{sivan}.\ 

This basic discrepancy was resolved by the work of Cohen \textit{et al}. 
\cite{cohen}. These authors solved the Hartree-Fock equations for a finite
disordered interacting system of charges on a lattice. This produced a
Gaussian shape for $P(x)$. The origin of this distribution is the Hartree
term in the total energy. The Hartree potential at any given site is a sum
of random variables, the charges at all the other sites weighted by their
inverse distance to the given site. Application of the central limit theorem
to this potential then yields the Gaussian form for the CELS. By making an
experimentally-guided estimate of the parameters in the model, Cohen \textit{%
et al. }also found agreement between theory and the experimental data of
Sivan et al. \cite{sivan} for the width of the distribution.

However, there remain unanswered questions. Some are experimental. As we
shall show in detail below, the most extensive data on $P(x)$ \cite{marcus}
show marked deviations from the Gaussian shape. In particular, there are
broad tails in the distribution. Furthermore, other experiments \cite%
{chandrasekhar} \ show some asymmetry in the distribution function,
indicating that the Gaussian is not universal.

There are also purely theoretical issues to be resolved. RMT is certainly
valid in regimes where the interaction is weak, as it is known to be correct
for non interacting systems. This means that there should be a crossover
regime from Wigner-Dyson to Gaussian statistics as the strength of the
interaction is increased and this has been seen in numerical studies \cite%
{alhassid}, \cite{gefen}. We analyze this in some more detail by finding the
crossover point in the presence of disorder with variable strength. \ Also,
it has been shown by Shklovskii \textit{et al.} \cite{boris} that
Wigner-Dyson statistics do not apply near the Fermi energy of Anderson
insulators. In this case the energy levels follow Poisson statistics. \
Thus, if the disorder dominates, we have yet a third kind of statistics,
again with crossovers that are in need of investigation. In this regard, it
is interesting to note recent work by Berkovits \textit{et al.,} who find a
crossover from Wigner-Dyson to Poisson statistics as a function of
interaction strength in the energy of the first excited state of dots \cite%
{berkovits}.\ Alhassid \textit{et al.} have seen the crossover from Gaussian
to Wigner-Dyson statistics, in a more generical model of a dot, valid at
small dimensionless resistance \cite{alhassid2}.

In this work, we take a synthetic approach to answer these experimental and
theoretical questions, in the hope of arriving at a global understanding of
the CELS of quantum dots. In Sec. \ref{sec:model}, we introduce a model that
includes interactions, disorder and hopping. We first examine the classical
limit of the model and then extend the arguments to the quantum case. The
qualitative results are summarized by means of a conjectured ``statistics
plot''. In Sec. \ref{sec:numerical}, we present the results of numerical
simulations to bolster the theoretical conclusions and make them somewhat
more quantitative. The final results are presented in Sec. \ref{sec:phase}.
Comparison to experiment is made in Sec. \ref{sec:experiments}, and our
conclusions are in Sec. \ref{sec:conclusion}.\\

\section{Model}\label{sec:model}

Our model of a dot is a system of $N$ interacting electrons on a disordered
lattice of $N_{s}$ sites: the Anderson Hamiltonian with long-range Coulomb
interactions:%
\begin{equation}
\mathcal{H}^{\prime }=\sum_{i}^{N_{s}}u_{i}^{\prime }n_{i}-t^{\prime}
\sum_{\left\langle ij\right\rangle }^{N_{s}}\left( c_{i}^{\dagger
}c_{j}+c_{j}^{\dagger }c_{i}\right) +e^{2}\sum_{i\neq j}\frac{n_{i}n_{j}}{%
\left\vert \vec{R}_{i}-\vec{R}_{j}\right\vert }.
\end{equation}%
In this equation, $i$ labels the sites of a finite square lattice. The sites
are located at $\vec{R}_{i}=(ma,na)$, where $m$ and $n$ are integers: $1\leq
m\leq L$ and $1\leq n\leq L.$ \ $\left\langle ij\right\rangle $ is a
nearest-neighbor pair, $n_{i}=c_{i}^{\dagger }c_{i}$ is the number operator,
and $\varepsilon _{i}^{\prime }$ are the site energies. The $u_{i}^{\prime }$
are drawn from a probability distribution $P(\varepsilon _{i}^{\prime })$ of
width $W$:

\begin{equation}
P(u_{i}^{\prime })=\left\{ 
\begin{array}{ccl}
1/W\,^{\prime }, & \textrm{for} & \left\vert u_{i}^{\prime }\right\vert
<W^{\prime }/2 \\ 
0, & \textrm{for} & \left\vert u_{i}^{\prime }\right\vert >W^{\prime }/2%
\end{array}%
\right. .
\end{equation}%
As we explain below, we will find the ground state of this Hamiltonian
numerically in the Hartree approximation. Since this approximation does not
respect the exclusion principle, double occupancy of sites is allowed. \ We
suppress this by adding an energy $U=4e^{2}/a$ for each doubly occupied
site. We will briefly consider the effect of varying $U$ below.

The model has three parameters $t^{\prime },W^{\prime },$ and $e^{2}/a$. \
Our main interest lies in the energy level statistics. These statistics can
only depend on two parameters, since one of the three can be scaled out. \
We choose $e^{2}/a$ as our energy unit and so define the dimensionless
quantities $\mathcal{H}=\mathcal{H}^{\prime }a/e^{2},$ $t=t^{\prime
}a/e^{2}, $ $\vec{r}_{i}=\vec{R}_{i}/a$, and $W=W^{\prime }a/e^{2}$, leading
to

\begin{equation}
\mathcal{H}=\sum_{i}^{N_{s}}u_{i}n_{i}-t\sum_{\left\langle ij\right\rangle
}^{N_{s}}\left( c_{i}^{\dagger }c_{j}+c_{j}^{\dagger }c_{i}\right)
+\sum_{i\neq j}\frac{n_{i}n_{j}}{\left\vert \vec{r}_{i}-\vec{r}%
_{j}\right\vert }  \label{eq:hamiltonian}
\end{equation}%
and%
\begin{equation}
P(u_{i})=\left\{ 
\begin{array}{ccc}
1/W & \textrm{for} & \left\vert u_{i}\right\vert <W/2 \\ 
0 & \textrm{for} & \left\vert u_{i}\right\vert >W/2%
\end{array}%
\right. .
\end{equation}

Our model of the dot is not the most general. However, we believe it is the
simplest one that combines the three essential features of the problem:
disorder, interaction, and hopping. It takes the simplest possible form for
the disorder, the simplest non-interacting band structure, and the simplest
long-range interaction. We shall have occasion to briefly investigate some
elaborations of the model such as different dot shapes and different
boundary conditions. It is generally believed that the Anderson model is
sufficiently general to capture all the qualitative features of many
physical properties having to do with disorder, localization being the prime
example. Given the intimate connection between localization and level
statistics, it seems plausible that this model is a good starting point for
our problem.

The chief difficulties in the numerical calculations are the necessities of
averaging over many realizations of the disorder and converging accurately
to the authentic ground state. In order to accomplish these two objectives,
we are forced to neglect the spin degree of freedom. This is undesirable,
particularly in view of suggestions that energy-level pairing might take
place, leading to bimodal distributions for the CELS. We only note that this
phenomenon is apparently absent in most experiments, and also in most of the
numerical work done previously. Since we will do calculations in the Hartree
approximation, we must also specify the onsite interaction, which is taken
as $U=4$. We discuss this choice further below.

To understand the level statistics of a particular dot, we model it by Eq. (%
\ref{eq:hamiltonian}) and then situate it on a plot of disorder versus
hopping strength: $W$ vs. $t$, and our task is to figure out the physics of
all the regions of the $W-t$ plane. We shall refer to this diagram as the
\textquotedblleft statistics plot\textquotedblright . The motivation for
plotting in this way is that the limiting regimes of the CELS can easily be
picked out. Let us discuss the theoretical expectations for these regimes in
turn.

The classical regime is defined by $t=0,$which is the vertical axis of the
statistics plot. Far out along this vertical axis of the statistics plot, $%
W\rightarrow \infty $ and the disorder is dominant, and we can neglect both
hopping and interactions. Each electron sits on a single site, and the sites 
$i$ are filled up in the order of $u_{i},$ from lowest to highest. \ The
ground state energy is given by%
\begin{equation}
E(N)=\sum_{i=1}^{N}u_{i},
\end{equation}%
where the $u_{i}$ are indexed such that $u_{1}<u_{2}<u_{3}<\cdot \cdot \cdot 
$ \ . Also%
\begin{equation}
\Delta _{2}=E_{G}(N+1)-2E_{G}(N)+E_{G}(N-1)=u_{N+1}-u_{N}.
\end{equation}%
This leads to Poisson statistics, essentially independent of the statistics
chosen for the $u_{i}$: 
\begin{eqnarray}  \label{eq:poisson}
P_{P}(\Delta _{2})=\left\{ 
\begin{array}{ccc}
e^{-\Delta _{2}/s}/s & \textrm{if} & \Delta _{2}>0 \\ 
0 & \textrm{if} & \Delta _{2}<0%
\end{array}%
\right.
\end{eqnarray}
and note that this is properly normalized. The capacitance $C$, which gives
a rigid shift $e^{2}/C$ in the distribution, is effectively infinite owing
to the absence of interactions in this limit. Because of the vanishing of $%
P_{P}$ at negative arguments, this is also an asymmetric distribution.

To understand the classical regime at weak disorder $W<<1$, it is necessary
to estimate the potential fluctuations. Consider a set of $N$ charges on the
lattice. In the absence of disorder ($W=0$), they will be distributed in
space in such a way as to make the entire dot into an equipotential surface,
up to atomic-scale graininess, which also gives graininess to the Coulomb
potential. If we add a small amount of disorder in an infinite system, we
expect that the site where the next charge goes to be determined entirely by
the graininess in the Coulomb potential and the weak randomness coming from
the very small disorder in site energies. Thus the width of the distribution 
$P(\Delta _{2},W),$ which we shall denote by $\sigma _{X}$ has a small
intercept on the $W$ axis and is linear in $W$: $\sigma _{X}(\Delta
_{2},W)=P_{0}+\beta W$, the first term coming from the graininess and the
second from the disorder. This process clearly leads to Gaussian statistics,
as the site energies are chosen at random and their sum is the total
energy.\ As detailed below, we find empirically that the atomic-scale
graininess is smaller than one might expect.

Thus at small disorder we have Gaussian CELS and at large disorder we have
Poisson CELS. At what value of $W$ \ does the crossover occur? \ 

Increasing the amount of disorder from small values we will find that some
charges, on entering the system, will end up one lattice constant away from
the site that is optimum for the Coulomb interaction. This costs an
electrostatic energy of order $\sim Ne^{2}a^{2}/L^{3},$ where $L$ is the
linear size of the lattice. The quadratic dependence on $a$ is due to the
fact that the potential energy is quadratic in the displacement, since the
charge is close to a potential minimum. This charge gains a site energy $%
\sim W.$ \ The number of displaced charges $N_{dis}$ is therefore of order $%
N_{dis}\sim NW/\left( Ne^{2}a^{2}/L^{3}\right) =WL^{3}/e^{2}a^{2}.$ \ Each
displaced charge creates a potential fluctuation at a test site that is of
order $e^{2}a/L^{2},$ since it is typically at a distance $\sim L$ from the
test site, but it has moved only by a distance $a.$ \ At the test site these
changes add randomly, giving rise to a potential at the test site that has a
normal distribution of width 
\begin{eqnarray*}
\sigma _{X} &\sim &\sqrt{N_{dis}}e^{2}a/L^{2}\sim \ \left(
WL^{3}/e^{2}a^{2}\right) ^{1/2}e^{2}a/L^{2} \\
&=&\left( We^{2}/L\right) ^{1/2}\sim \left( \frac{W}{e^{2}/C}\right) ^{1/2}%
\frac{e^{2}}{C} \\
&=&\left( \frac{W}{E_{c}}\right) ^{1/2}E_{c}.
\end{eqnarray*}%
Here $E_{c}$ is the charging energy $e^{2}/C\approx e^{2}/L.$ \ These
considerations hold for infinite systems. In finite systems, a substantial
fraction of the charges will be at or near the boundary of the system. \
Moving these charges costs a much larger amount of Coulomb energy $\sim
e^{2}/a.$ \ This is a very important effect in our calculations because of
the small lattice sizes and small number of charges. In such a system the
surface effects increase the size of the regime of small $W$ where the
disorder cannot affect the position of the charges. Increasing $W$ in this
regime does not change the Coulomb energy and the width of the distribution
comes entirely from the disorder energy. The site energies are drawn
randomly from the uniform (or other) distribution an hence the differences
are normally distributed. Hence we expect a width proportional to $W$ for
very small $W$ and to $W^{1/2}$ for slightly larger $W.$ \ Let us denote the
value of $W$ where this first crossover takes place as $W_{cr}^{(1)}.$ \ For 
$W<W_{cr}^{(1)}$ disorder dominates, while the Coulomb energy takes over for 
$W>W_{cr}^{(1)}.$ \ Thus we expect $W_{cr}^{(1)\prime }\sim e^{2}/a$ which
in dimensionless units gives $W_{cr}^{(1)}\sim 1.$ \ \ 

However, this is not the crossover to genuine Poisson statistics, a
one-sided exponential form for $P(\Delta _{2}).$ \ That crossover only
occurs when the $N+1st$ charge, added into a background of randomly-placed
charges, and with a choice of order $N_{s}$ sites, must always choose the
one which has the lowest site energy rather the one with lowest Coulomb
energy. This will occur when $W^{\prime }/N_{s}$ is comparable to $e^{2}/a. $
\ This yields $W_{cr}^{(2)}\sim N_{s}>W_{cr}^{(1)}\sim 1$ as the crossover
to Poisson statistics.

What about the intermediate regime $W_{cr}^{(1)}<W<W_{cr}^{(2)}$? \ In our
numerical studies, as we shall see below, in this intermediate regime we
find a symmetric distribution with broader tails than one has in a Gaussian.
\ We suggest the following rather speculative explanation. In this general
case, the Coulomb and disorder energies both contribute. We can consider the
distribution of $\Delta _{1}(N)=E_{G}(N+1)-E_{G}(N)-Ne^{2}/2C$ directly. \
This is a random variable whose instances are indexed by $N.$ \ Its
distribution is centered on $0$ by the definition of $C$. If there is no
correlation between $\Delta _{1}(N)$ and $N$ itself, then the second
difference $\Delta _{2}=\Delta _{1}(N)-\Delta _{1}(N-1)$ is the difference
of two values drawn from this distribution at random and $P(\Delta _{2})$
will be Gaussian. However, in the limit of strong disorder, this is clearly
not so: $E_{G}(N+1)-E_{G}(N)=u_{N+1},$ which is an increasing function of $N.
$ \ When disorder is somewhat weaker, and minor charge redistribution is
allowed, we may still expect that the particle number is changed only by a
small amount, then there can be regions of energy (or particle number),
where successive (in energy) instances of the random variable $%
E_{G}(N+1)-E_{G}(N)-Ne^{2}/2C$ correspond to changes in particle number by
only one particle. Then $\Delta _{1}$ satisfies Poisson statistics as in Eq.
(\ref{eq:poisson}) and $\Delta _{2}$ satisfies%
\begin{eqnarray*}
P_{E}(\Delta _{2}) &=&\frac{1}{s^{2}}\int_{0}^{\infty }d\Delta
_{1}e^{-\Delta _{1}/s}\int_{0}^{\infty }d\Delta _{1}^{\prime }e^{-\Delta
_{1}^{\prime }/s}\delta \left[ \Delta _{2}-\left( \Delta _{1}-\Delta
_{1}^{\prime }\right) \right] \\
&=&\frac{1}{2s}\exp (-\left\vert \Delta _{2}\right\vert /s),
\end{eqnarray*}%
which is a two-sided exponential, a symmetric distribution. This is a
reasonably good representation of what we find in the numerics.

Along the other axis $W=0$ the system is fully quantum and a good starting
point to understand the CELS\ is the Hartree approximation. For $N$
particles, the Hartree Hamiltonian, in reduced units, is 
\[
\mathcal{H}_{H}(N)=-t\sum_{\left\langle ij\right\rangle }^{N_{s}}\left(
c_{i}^{\dagger }c_{j}+c_{j}^{\dagger }c_{i}\right) +\left(
\sum_{i}^{N_{s}}u_{i}+\sum_{i\neq j}\frac{n_{j}}{\left\vert \vec{r}_{i}-\vec{%
r}_{j}\right\vert }\right) c_{i}^{\dagger }c_{i}. 
\]%
It has eigenfunctions $\psi _{\alpha }(N,i)c_{i}^{\dagger }\left\vert
0\right\rangle $ and eigenvalues $\varepsilon (N,M).$ \ The first index $N$
indicates the total number of particles, and the second index labels the
eigenvalues in increasing order. The density%
\[
n_{j}(N)=\sum_{\alpha =\textrm{occupied}}^{N}\left\vert \psi _{\alpha
}(N,i)\right\vert ^{2} 
\]%
must be calculated self-consistently. The ground state energy is%
\[
E_{G}(N)=\sum_{M=1}^{N}\varepsilon (N,M)-\frac{1}{2}\sum_{i\neq j}\frac{%
n_{i}(N)n_{j}(N)}{\left\vert \vec{r}_{i}-\vec{r}_{j}\right\vert } 
\]%
\ This approximation has long been used to calculate ionization energies in
atoms and molecules. These energies are analogous to our charging energies.
This is usually done by means of Koopman's relation 
\[
E_{G}(N+1)-E_{G}(N)=\varepsilon (N,N+1). 
\]%
Cohen \textit{et al. }\cite{cohen} point out that this implies 
\[
\Delta _{2}=E_{G}(N+1)-2E_{G}(N)+E_{G}(N-1)=\varepsilon (N,N+1)-\varepsilon
(N-1,N). 
\]
In contrast, a particle-hole excitation corresponds to $\varepsilon
(N,N+1)-\varepsilon (N,N).$ \ Since the two eigenvalues $\varepsilon
(N,N+1)-\varepsilon (N,N)$ come from a single Hamiltonian, we expect and
Cohen \textit{et al}. find that the differences of this kind follow
Wigner-Dyson-type statistics. But the quantities of interest to us, $%
\varepsilon (N,N+1)$ and $\varepsilon (N-1,N),$ are drawn from different
Hamiltonians and therefore from separate probability distributions. We can
get Gaussian statistics for their difference $\Delta _{2}$ because of the
Gaussian character of the fluctuations in the Coulomb potential \cite%
{alhassid}, \cite{cohen}. Note that as the interactions become weak, $%
\varepsilon (N,M)$ becomes independent of $N.$\ 

This discussion allows to understand the horizontal axis of the statistics
plot. Far out along the axis, the hopping term dominates and disorder and
interaction can be neglected. For a highly symmetric lattice such as we
shall consider in our numerical work, we get nonuniversal results very close
to the axis for the CELS. This is due to symmetry-related degeneracies that
are of no interest for the present study. Fortunately, this non-universal
regime is very narrow, since a small amount of disorder lifts the
degeneracies. Irregular dot shapes would presumably have the same effect. We
shall therefore ignore the $t$-axis itself, since it is unlikely to apply to
real dots. Just off the axis, but far out along it, interactions are
unimportant and we find Wigner-Dyson statistics, with the distribution
function given by \ref{eq:wdstat}.

There is a crossover to Gaussian statistics when the interactions become
more important near the origin in the statistics plot. This crossover is
governed by the density parameter $r_{s}=1/\sqrt{4\pi n_{s}a_{B}^{\ast }}$,
where $n_{s}$ is the area electron density and $a_{B}^{\ast }=\hbar
^{2}/m^{\ast }e^{2}$ is the effective Bohr radius. In terms of our
parameters,%
\[
r_{s}=\sqrt{\frac{N}{16\pi }}\frac{1}{t}. 
\]%
The crossover is expected when $r_{s}>1.$ \ This has been found in several
studies \textit{\ }\cite{cohen}, \cite{alhassid}.

In the two-dimensional $W-t$ plane the question is how disorder destroys the
Gaussian statistics as $t$ increases. At $t=0$ $(r_{s}=\infty ),$ the
crossover to non-Gaussian occurs by definition at $W_{cr}^{(1)}.$ \ What is $%
W_{cr}^{(1)}(t)?$ \ We determine this numerically, but we expect that it is
an increasing function. Below the crossover, the choice of the state to be
filled by the $N+1st$ particle is determined mainly by the Coulomb
interaction. This will be easier if the states are spread out than when they
are localized on sites, as in the $t\rightarrow 0$ limit.

A third region that can be characterized on the statistics plot is far out
along any line through the origin with finite slope $d=W/t$ , for then the
interaction may be neglected and the behavior of the system is determined by
the dimensionless ratio $d$ that characterizes the now independent
electrons. If we imagine traveling in clockwise fashion around a circle
centered at the origin with very large radius we expect a crossover from
Poisson CELS at large $d$ to Wigner-Dyson CELS at small $d$.

On the statistics plot, the origin is the strong-coupling point. Since
Gaussian CELS\ result from interactions, we expect a Gaussian regime very
near the origin, but classical crystallization must also occur near this
point, and the effects of geometry have been investigated by Koulakov and
Shklovskii \cite{koulakov}. It is unclear how crystallization influences the
CELS. We shall not be concerned, except in passing, with crystallization
issues in this paper. \ 

\begin{figure}[tbp]
\centerline{ 
\begin{tabular}{cc}
\includegraphics[
  scale=0.55]{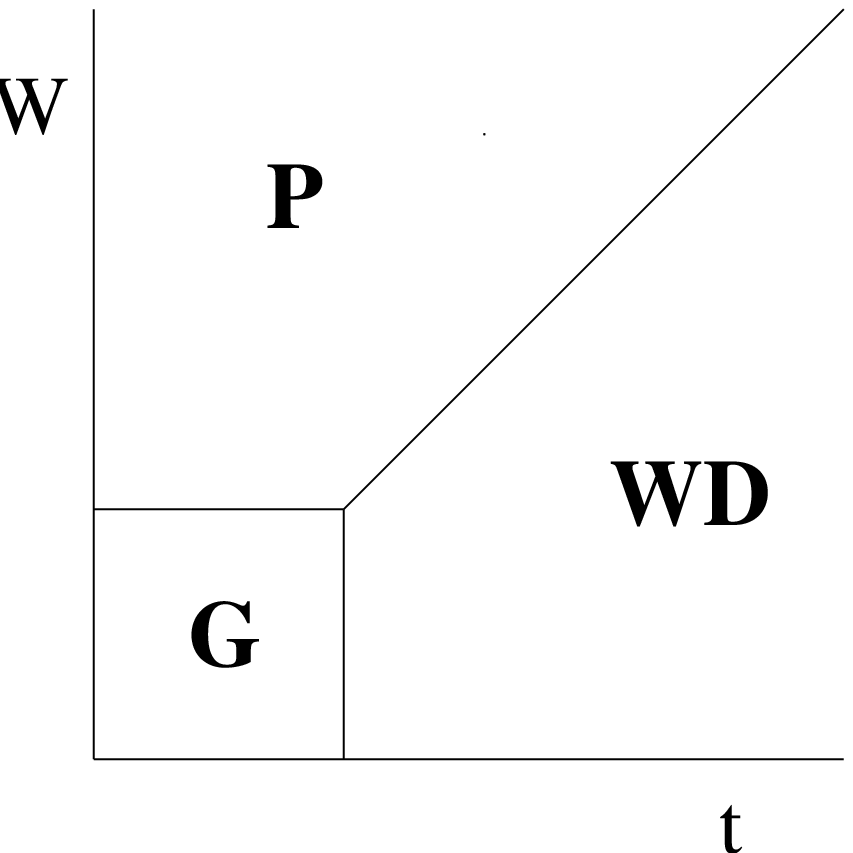}~~~~~~~~~~~~&
\includegraphics[
  scale=0.55]{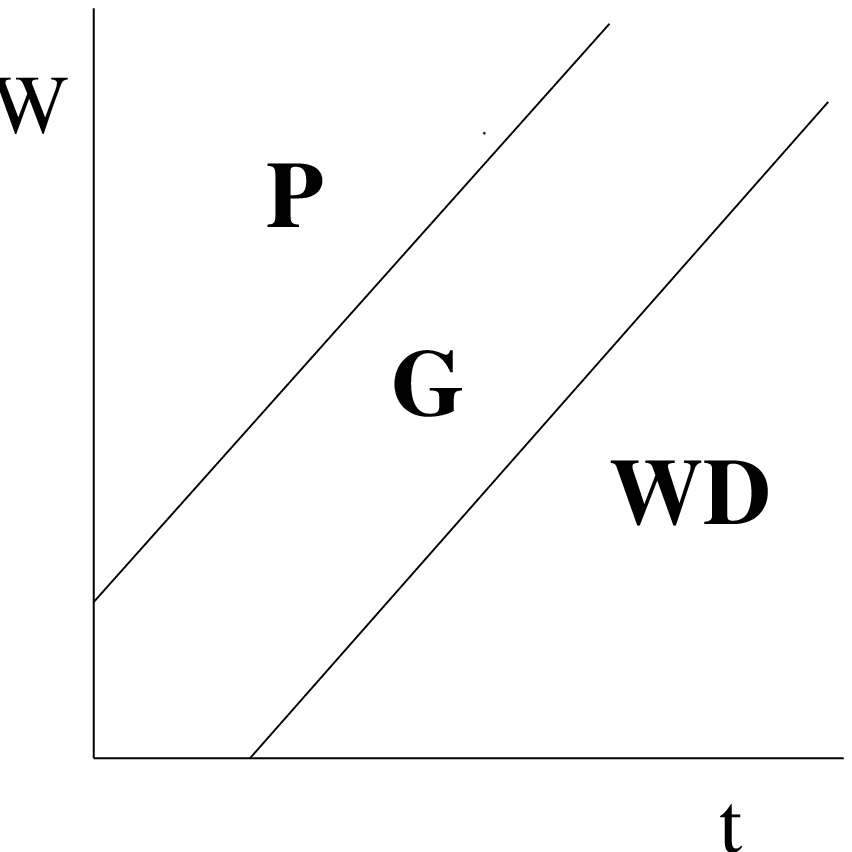}\tabularnewline
(a)~~~~~~~~~~~~&
(b)\tabularnewline
\tabularnewline
\includegraphics[
  scale=0.55]{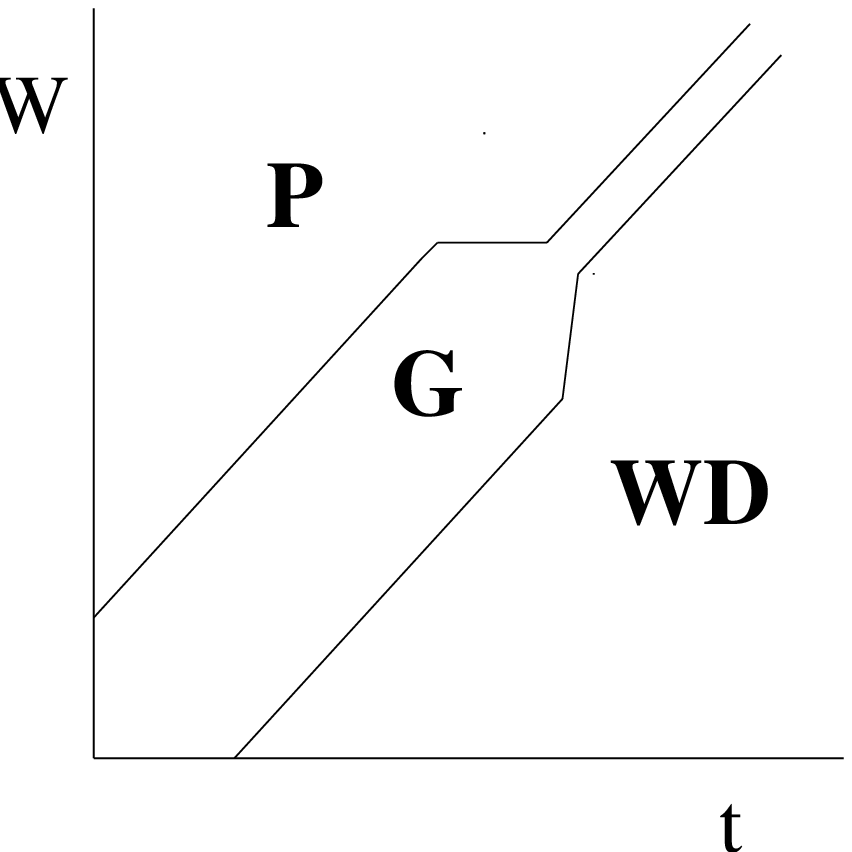}~~~~~~~~~~~~&
\includegraphics[
  scale=0.55]{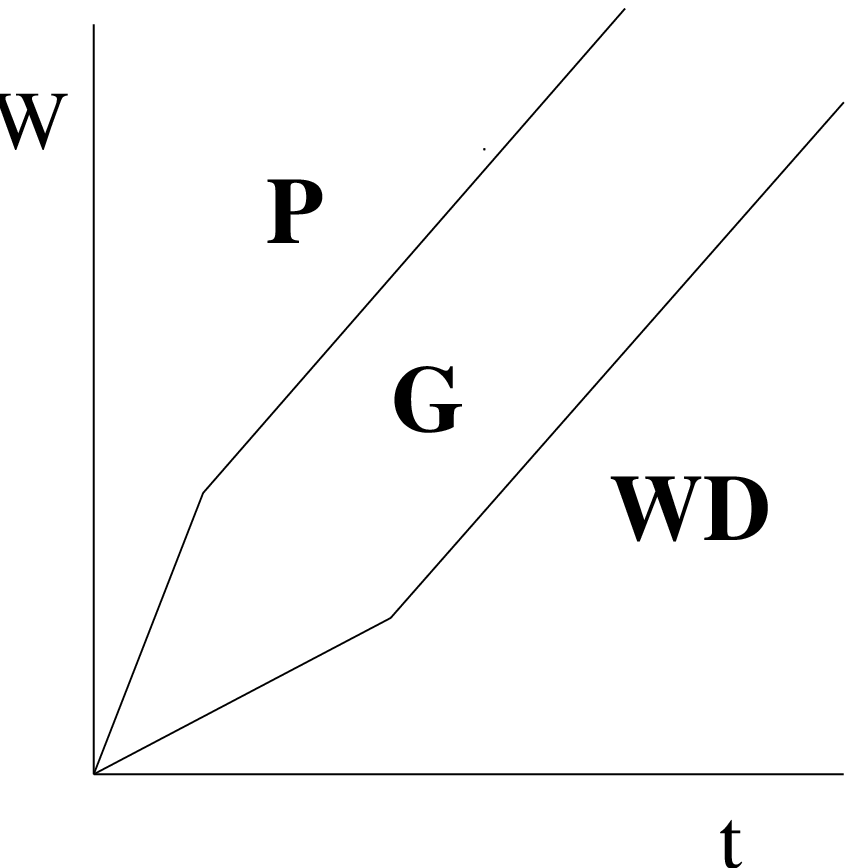}\tabularnewline
(c)~~~~~~~~~~~~&
(d)\tabularnewline
\end{tabular}
} 
\caption{Possible topologies for the statistics plot. $W$ is a measure of the
disorder, while $t$ is the dimensionless hopping strength. P stands for
Poisson charging energy level statistics (CELS), G for Gaussian CELS, and WD
for Wigner-Dyson CELS.}
\end{figure}
These considerations of limits do not determine the topology of the
statistics plot completely. Some possibilities are shown in Fig. 1. 
\ In 1(a), there are critical values of $W$ and $t$ beyond which Gaussian
CELS\ cease entirely, while in 1(b), and 1(c), this is not the case.\ \
There is a critical value of $W/t$ characteristic of the noninteracting
problem that is common to all three possibilities. 1(b) and 1(c) are
distinguished by the width of the Gaussian region as the interaction
strength becomes weaker (for varying $W$ at fixed $W/t)$. This width may or
may not vanish. In 1(a), 1(b), and 1(c) there can be transitions from
Poisson or Wigner-Dyson ELS to Gaussian ELS\ as only the interaction is
changed, i..e., when one starts at an arbitrary point and moves toward the
origin along a straight line. In 1(d), this cannot occur. A\ long-term goal
\ would be to decide between these various topologies.

The lines on the plots of course do not separate distinct phases, and we
must not interpret the statistics plot as a phase diagram, though the
analogy is in some ways useful. As we are dealing with a finite system, we
would only expect crossovers even in classical thermodynamics. Here there is
additional physics that further smooths the transitions. For example, we
have said that we expect Poisson CELS when the states are all localized. \
However, it is known (at least in the noninteracting case) that all states
are localized in two dimensions. However, the localization length generally
depends on energy as well as disorder strength and other parameters. If we
probe the CELS when the Fermi energy is such that the localization length is
long compared to the size of the system, we have the possibility of
Wigner-Dyson CELS. Hence the statistics are not necessarily independent of $N
$, the number of particles, even in the noninteracting case. When
interactions are added, and the density changes with $N$, this conclusion is
strengthened even more. Our main interest, however, lies in the topology of
the plot, and those quantitative features that are reasonably robust, to be
discussed further below. We expect some of these features to be independent
of $N$ or to vary weakly with $N$.

The virtues of attempting to understand dot CELS through the statistics plot
are several. The first is that it offers a global picture of CELS, which
summarizes all possibilities succinctly. The plot can serve as a diagnostic
tool in the experimental investigation of a specific dot or type of dot:
where does the dot lie in the $W-t$ plane? \ It gives a way of connecting
the classical and quantum cases in a continuous fashion. Since classical
electrostatic effects certainly play some role in dot physics, this is
important. Finally, in fitting data for $P(\Delta _{2})$, we believe it is
essential to have interpolation formulas that combine the three types of
CELS, and the statistics plot gives us guidance as to how to accomplish this.

\section{Numerical Calculations}\label{sec:numerical}

\subsection{Classical region}

\begin{figure}[tbp]
\centerline{
\begin{tabular}{cc}
\includegraphics[scale=.7]{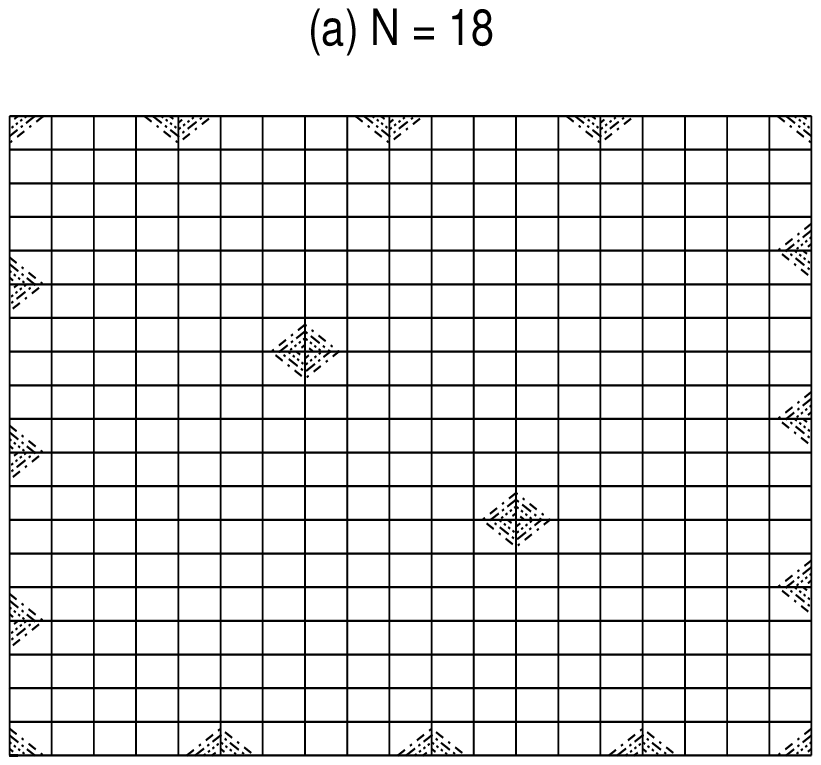}~~~
 &\includegraphics[scale=.7]{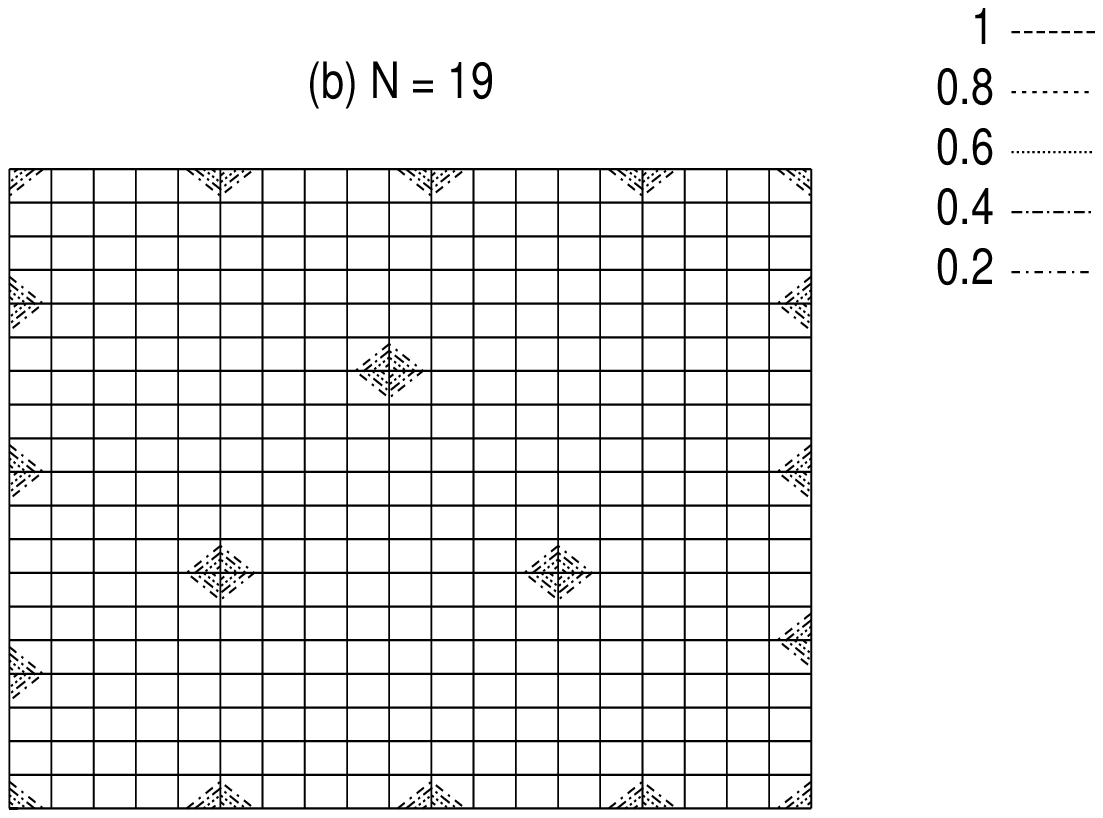}
\tabularnewline
\end{tabular}
} 
\caption{The ground state charge density distributions for (a) $N=18$ and
(b) $N=19$. Note that there is a substantial rearrangement of charge when
 the particle number is changed.  However, the energy change is much smaller
 because of high near-degeneracy among the low-lying configurations.}
\end{figure}
\begin{figure}[tbp]
\centerline{\includegraphics[scale=.9]{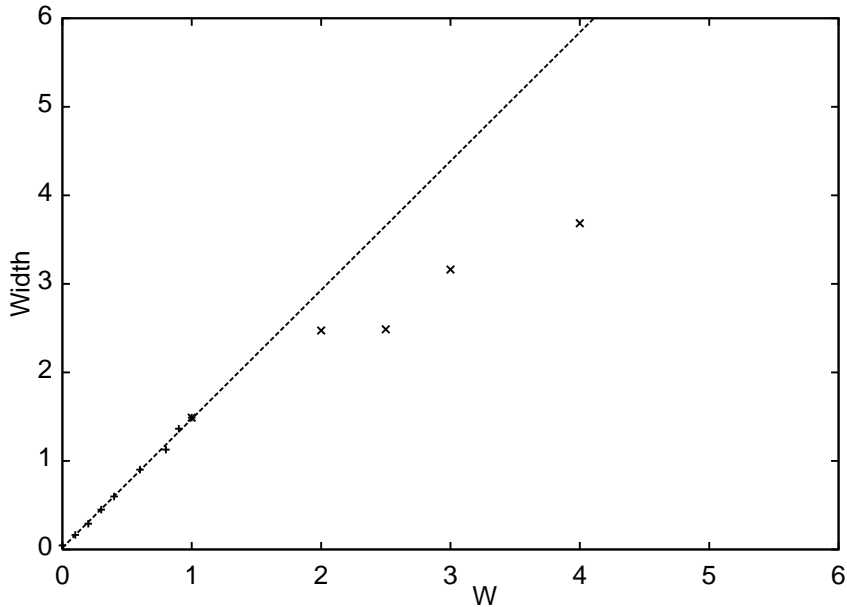}}
\caption{The root-mean-square width of the peak spacing distribution vs. $W$%
, a measure of the disorder parameter. The width increases linearly with the
disorder at small $W$ then crosses over to a more slowly growing function,
approximately proportional to $\protect\sqrt{W}$.}
\end{figure}

This section is devoted to numerical calculations of the CELS in the the
classical limit of the model defined by Eq. (\ref{eq:hamiltonian}). We set $%
t=0$, and restrict our attention to the vertical axis of the statistic plot.
The ground state energy of the $N$-particle system may be written as%
\begin{equation}
E(N)=\sum_{i=1}^{N_{s}}n_{i}u_{i}+\frac{1}{2}\sum_{i\neq j}^{N_{s}}\frac{%
n_{i}n_{j}}{\left\vert \vec{r}_{i}-\vec{r}_{j}\right\vert },
\label{eq:classical}
\end{equation}%
where the occupation numbers $n_{i}=0,1$ are chosen to minimize $E(N)$
subject to the constraint $\Sigma _{i}n_{i}=N$. This is a type of disordered
Ising model, and it presents a very difficult optimization problem in finite
geometries.

The method of choice for finding the ground state is the genetic algorithm,
whose application to this problem we now describe. We wish to minimize the
expression in Eq. (\ref{eq:classical}) with respect to the $n_{i}.$ \ Our
particular implementation is as follows. We first randomly choose a
particular realization of the disorder. Then we choose 10 candidate
solutions. One of these is the solution to the non-interacting problem,
given by occupying the sites having the lowest site energies. The rest are
chosen randomly. Each solution is relaxed by local movements. That is, each
particle is allowed to move to a nearby site if that lowers the energy, and
this is continued until no further movements are made, so that a local
minimum is found. (This part of the algorithm is 'greedy'.) \ The energies
of these 10 relaxed solutions are evaluated, and only the 5 of lowest energy
are chosen to survive. Exceptions are made to this rule when two or more
configurations are very similar in energy. In this case one or more is
discarded to preserve genetic diversity. The surviving configurations are
mated with each other by combining the top half ($n>L/2$) of one
configuration with the bottom half ($n\leq L/2$) of another configuration. \
Minor exceptions to the definition of top and bottom half must be allowed so
as to conserve particle number in the mating process. This produces the
second generation, and the process of relaxation, evaluation, selection and
mating is iterated. In general, we found that convergence was reached after
about 20 generations of this evolutionary process. Since there is disorder
it is necessary to average over many realizations, and this is what makes
the computations time-consuming. We found that 50 realizations produced
convergent results. In addition, if $P(\Delta _{2})$ is desired, averaging
over particle number $N$ is needed. We averaged $N\,$\ from $5$ to $40$ on a 
$20\times 20$ lattice. This allows us to compute $C,$ the capacitance, since
we define $e^{2}/C$ as the average value of $\Delta _{2}.$ \ It comes out to
be about $C=8a$ in our model.

To understand the averaging over $N,$ recall that $r_{s}=\sqrt{N/16\pi t^{2}}%
.$ \ So in a single point on our graph, $r_{s}$ is averaged over a range of $%
\left( 0.3-0.9\right) /t.$ \ This does not appear to introduce serious
errors: we checked in selected cases whether there was a strong $N-$%
dependence in the distribution function by doing subaverages over different
ranges of $N$ with other parameters fixed. These dependencies appeared to be
small.

It is of some interest to see the explicit results for the charge densities
in the ordered system. As the particle number increases, substantial
rearrangement of the charge takes place. For small numbers of particles,
these changes are clearly shape-dependent and some of the ground states in
the square are shown in Fig. 2. For larger numbers of particles, the
triangular lattice forms and the configurations can be described in terms of
this lattice and its standard defects. The defects form in order to fit the
lattice into the boundary \cite{koulakov}. Thus there are classical shell
effects that might be expected to contribute to the CELS. However, we
generally found that the differences in Coulomb energies between competing
configurations was quite small. Hence, as a particle is added, the
rearrangements of charge can be very significant, particularly at small $N.$
\ However, this does not give rise to anomalously large fluctuations in $%
\Delta _{2}.$ \ Even a very small amount of disorder or hopping is more
important than the classical shell effects, meaning that they do not have
much influence on the shape of the statistics plot as a whole.

The width of the distribution at small $W$ is plotted in Fig. 3. We see that
the linear behavior at very small $W$ does indeed appear to cross over to
square-root behavior, as predicted theoretically above. However, the
numerical data are noisy and it is not possible to extract an exponent with
any quantitative precision. The very small value of the intercept on the $%
\sigma _{X}$ axis is the basis of the statement that the shell effects that
would produce graininess in the Coulomb potential, are quite small. The
first crossover takes place at about $W_{cr}^{(1)}\sim 1,$ as expected. \
This coincides more or less with the crossover from a Gaussian $P(\Delta
_{2})$ to a two-sided exponential $P(\Delta _{2}).$ \ In Fig. 4. we plot the
numerically determined $P(\Delta _{2})$ as a function of $W.$ \ The
crossover is unmistakable. 
\begin{figure}[tbp]
\begin{tabular}{cc}
{\footnotesize P($\Delta_{2}$)}\begin{tabular}{c}
\includegraphics[scale=0.5]{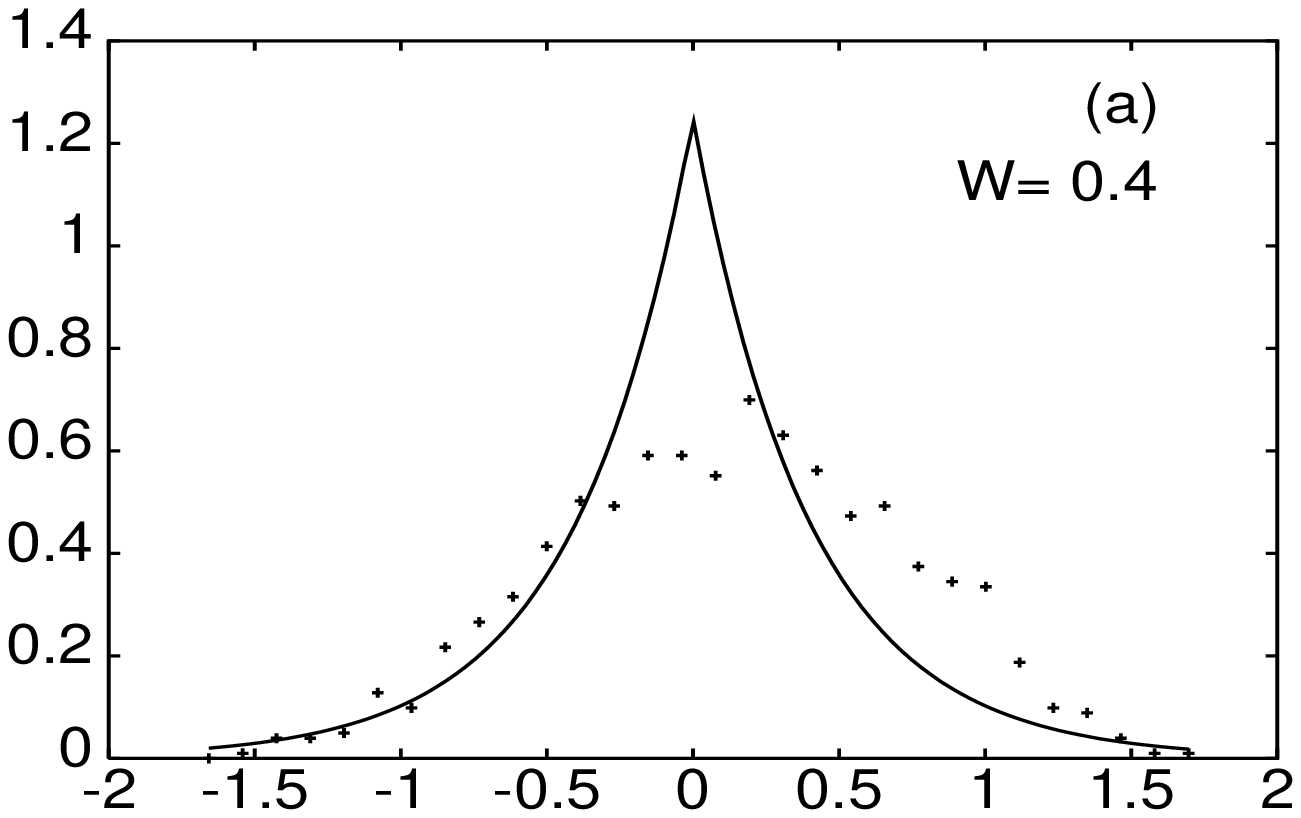}\tabularnewline
{\footnotesize $\Delta_{2}$}\tabularnewline
\end{tabular}&
{\footnotesize P($\Delta_{2}$)}\begin{tabular}{c}
\includegraphics[scale=0.5]{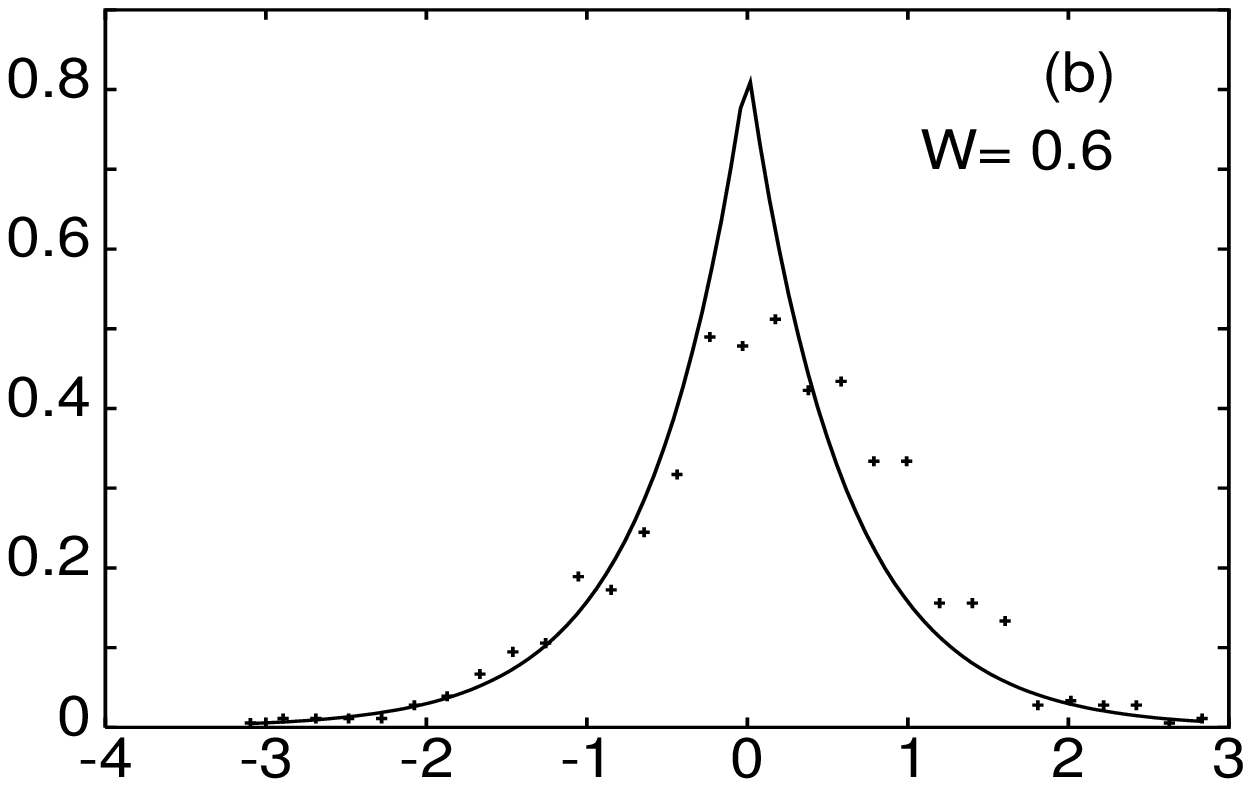}\tabularnewline
{\footnotesize $\Delta_{2}$}\tabularnewline
\end{tabular}\tabularnewline
{\footnotesize P($\Delta_{2}$)}\begin{tabular}{c}
\includegraphics[scale=0.5]{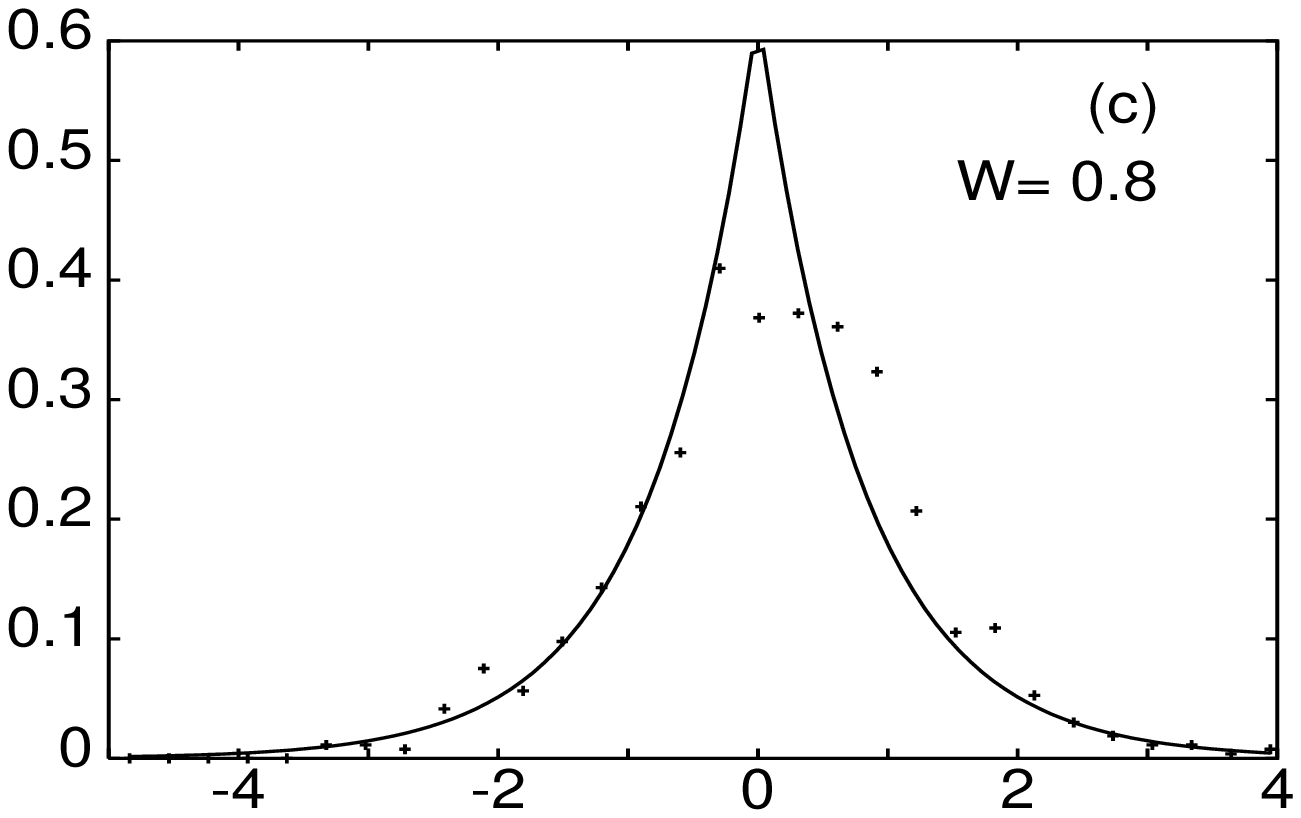}\tabularnewline
{\footnotesize $\Delta_{2}$}\tabularnewline
\end{tabular}&
{\footnotesize P($\Delta_{2}$)}\begin{tabular}{c}
\includegraphics[scale=0.5]{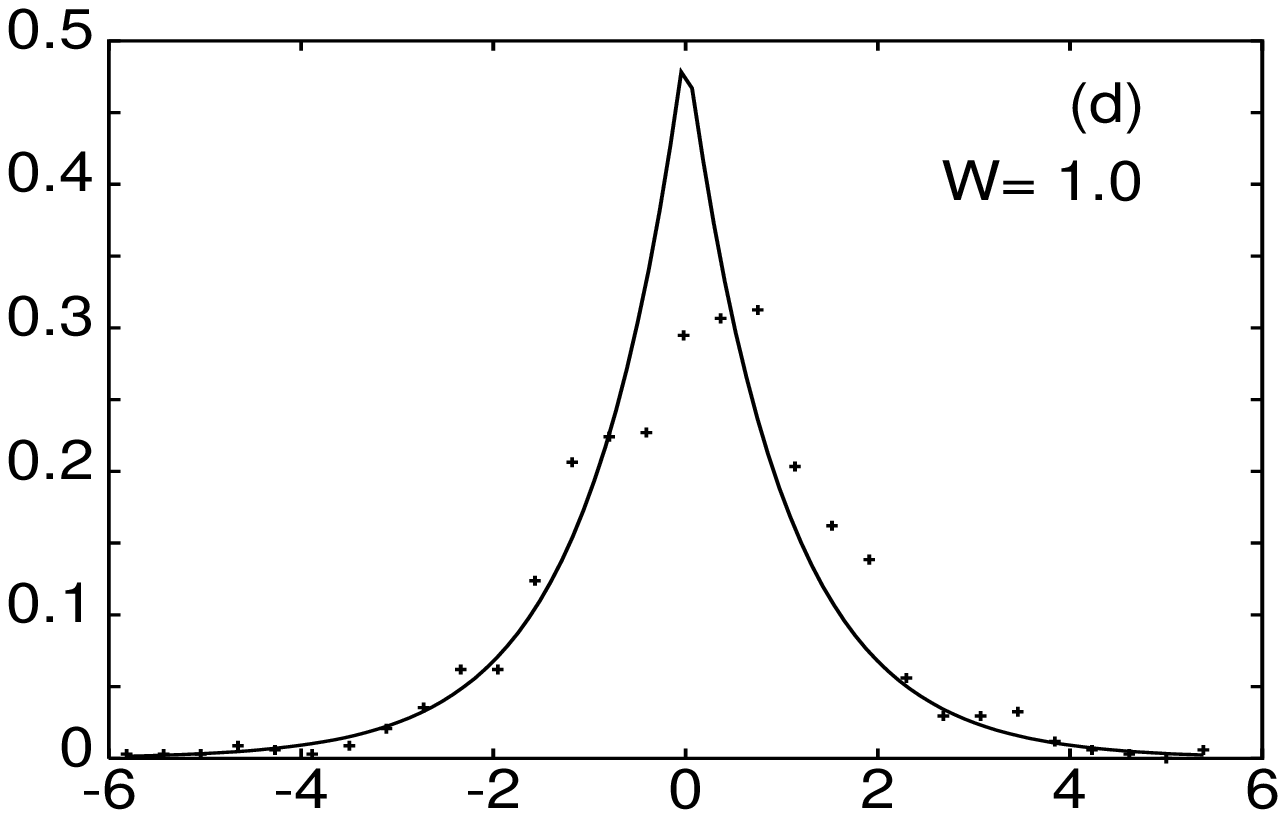}\tabularnewline
{\footnotesize $\Delta_{2}$}\tabularnewline
\end{tabular}\tabularnewline
{\footnotesize P($\Delta_{2}$)}\begin{tabular}{c}
\includegraphics[scale=0.5]{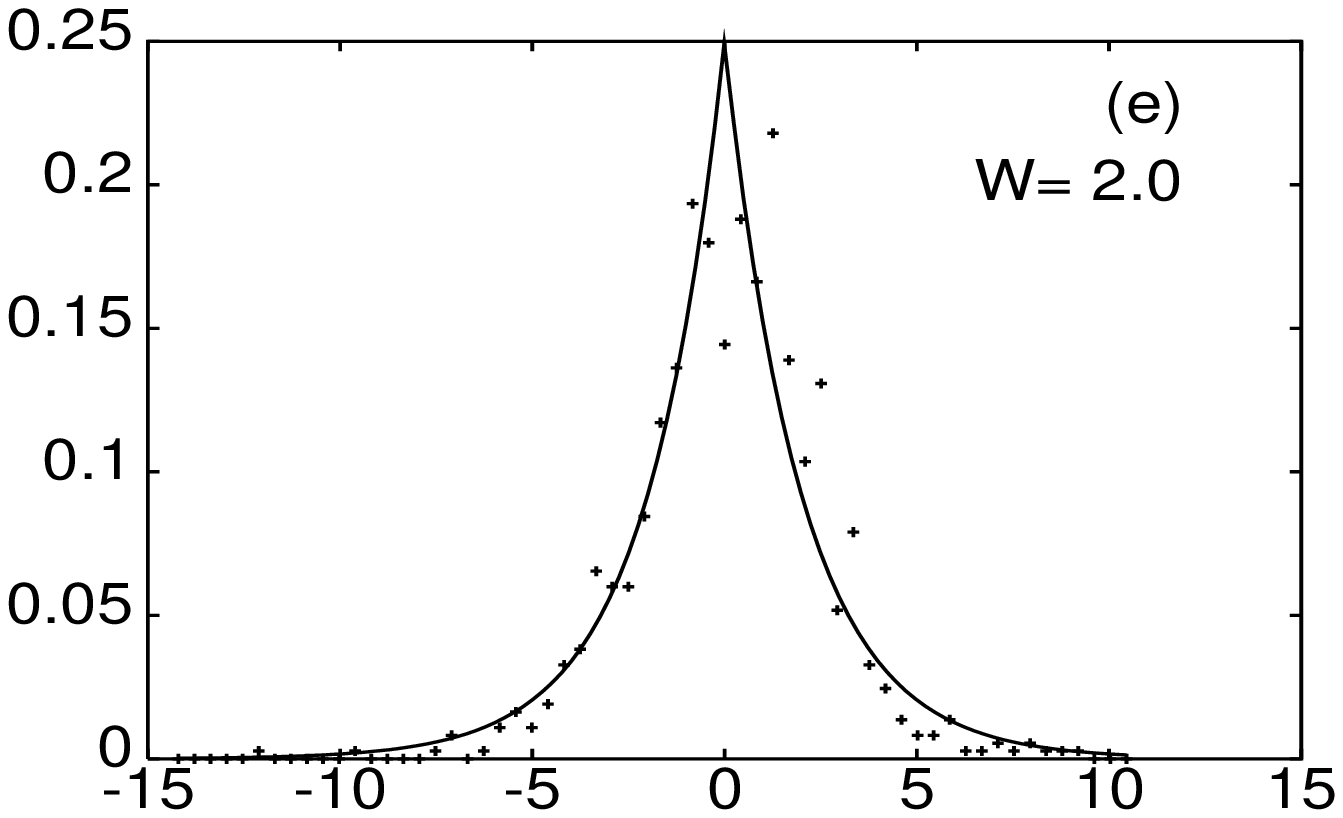}\tabularnewline
{\footnotesize $\Delta_{2}$}\tabularnewline
\end{tabular}&
{\footnotesize P($\Delta_{2}$)}\begin{tabular}{c}
\includegraphics[scale=0.5]{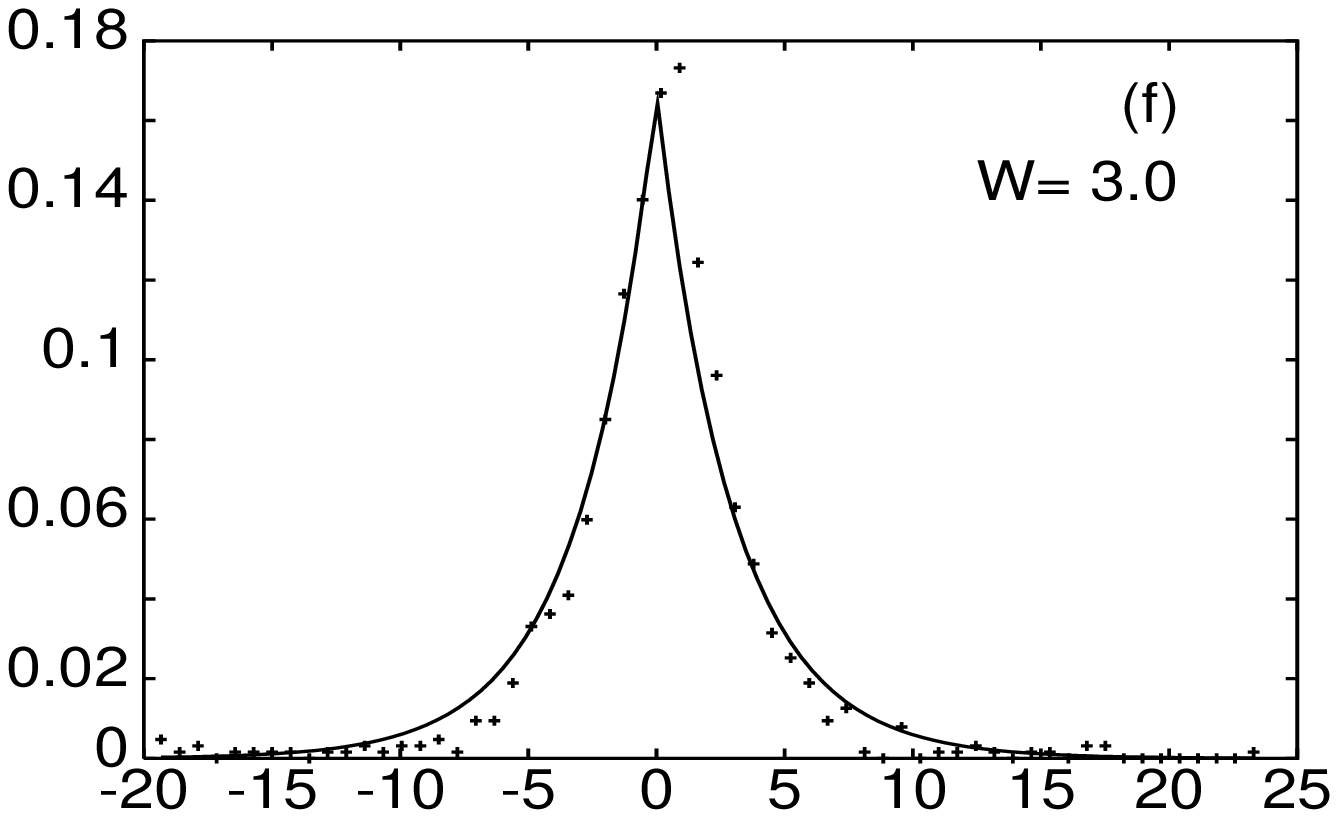}\tabularnewline
{\footnotesize $\Delta_{2}$}\tabularnewline
\end{tabular}\tabularnewline
\end{tabular}

\caption{Comparison of numerical results and the two-sided exponential
function for the CELS function $P(\Delta _{2}).$ $W$ is a measure of the
disorder. \ At small $W$, $P(\Delta _{2})$ is approximately Gaussian, but
for $W>2$ the two-sided exponential fit is better, suggesting the start of a
crossover.}
\end{figure}

The further crossover at $W_{cr}^{(2)}$ to Poisson statistics was located by
fitting $P(\Delta _{2})$ to the Poisson distribution Eq.(\ref{eq:poisson})
and a symmetrical Gaussian distribution:%
\[
P_{G}(\Delta _{2})=\left( 2\pi \sigma \right) ^{-1/2}\exp \left[ -\left(
\Delta _{2}-e^{2}/C\right) ^{2}/2\sigma ^{2}\right] .
\]%
\ Note that the Gaussian has two parameters, as opposed to the single
parameter in the Poisson expression in Eq. (\ref{eq:poisson}). The goodness
of fit is determined by the usual\ $\chi ^{2}$, the mean-square deviation of
the numerical points from the theoretical distributions. Some representative
fits are shown in Fig. 5. In order to compare the two fits as a function of $%
W$, we normalize the $\chi ^{2}$ as follows:%
\[
\gamma _{P}=\frac{\sum_{i}\left[ P(\Delta _{2i})-P_{P}(\Delta _{2i})\right]
^{2}}{\sum_{i}\left[ P_{G}(\Delta _{2i})-P_{P}(\Delta _{2i})\right] ^{2}}
\]%
and%
\[
\gamma _{G}=\frac{\sum_{i}\left[ P(\Delta _{2i})-P_{G}(\Delta _{2i})\right]
^{2}}{\sum_{i}\left[ P_{G}(\Delta _{2i})-P_{P}(\Delta _{2i})\right] ^{2}}.
\]%
These goodness-of-fit parameters are plotted in Fig. 5 in the regime of
large $W$. When $\gamma _{P}<(>)\gamma _{G},$ then Poisson (Gaussian)
statistics best describe the distribution. The crossover happens at about $%
W_{cr}^{(2)}=0.75N_{s}$, in reasonable agreement with the considerations of
the previous section. For values of $W$ which exceed this, the statistics
are Poisson. 
\begin{figure}[tbp]
\centerline{\includegraphics[scale=.95,angle=-90]{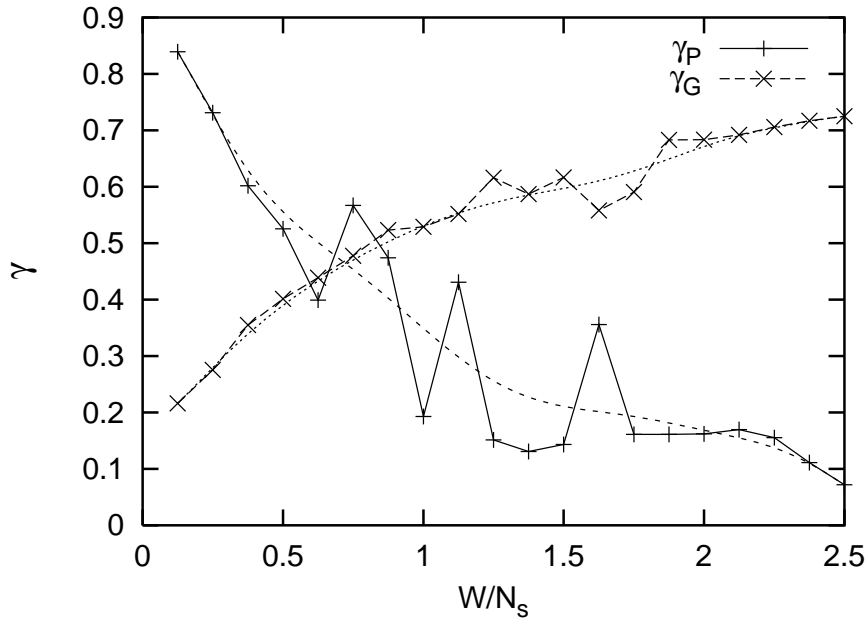}}
\caption{Goodness-of-fit for the case $t=0$. $\protect\gamma $ is small when
the fit is good.  $\protect\gamma _{G}$ and $\protect\gamma _{P}$ refer to
the Gaussian and Poisson distributions, respectively. \ When the disorder is
small the distribution is closer to Gaussian. As the strength of disorder
increases the Poisson distribution fits better. The +'s and $\times $'s are
numerical data, whereas the solid and dashed lines are smoothed data. The
crossover occurs roughly at $W/N_{s}=0.75$.}
\label{fig:poisson}
\end{figure}

It would be of considerable interst to understand the crossover from the
two-sided exponential to the one-sided exponential (Poisson) distribution.
However, we have not been able to characterize this crossover very
precisely. There are continuous ways to go between the one- and two-sided
distributions, such as simply having separate prefactors for the two sides
of the distribution. The numerics are consistent with this kind of
crossover, but are not sufficient to verify it in detail.

\subsection{Quantum Region}

This section is devoted to the quantum case, which is defined by the full
Hamiltonian of Eq. (\ref{eq:hamiltonian}). Our approach is to approximate
the solution by the Hartree approximation. The justification for this is
that Fock terms, not having a definite sign, tend to give a much smaller
contribution to the single-particle energies than the Hartree terms. This
has been confirmed numerically by Cohen \textit{et al}. \cite{cohen}. A
secondary justification is that doing parameter studies and averages over
realizations would not be computationally feasible if the complicated Fock
terms were retained.

$P(\Delta _{2})$ is determined by finding the ground state energies of the
operator in Eq. (\ref{eq:hamiltonian}) in the Hartree approximation. This is
done numerically on a finite square lattice. As in the classical case, we
take the size of the lattice to be $20\times 20$ and vary the particle
number between $5$ and $50$ and average over the particle number and $50$
realizations of the disorder to find $P(\Delta _{2})$. As stated above, we
consider only spinless fermions. Our calculation is self-consistent, in the
sense that the Hartree potential is iterated to convergence. The accuracy of
the ground state energy, as judged by the change in the last step of the
iteration, is typically one part in $10^{-3}$. If we take a more accurate
measure of the error, which is the difference between the final energies for
different starting configurations, we typically find an error of one part in 
$10^{-2}$.\ \ \ As a fraction of the width of $P(\Delta _{2})$, this is
normally a few percent. Thus we do not believe numerical errors
substantially influence the shape of the calculated distribution function.
In order to maintain this level of accuracy, we found that the calculations
needed to be restricted to the regime $t>0.3.$ \ For smaller $t$ values, the
convergence becomes slow and the accuracy quickly worsens. \ This is what
necessitates the special methods (such as the genetic algorithm) for the
classical case. It is interesting that the inclusion of quantum effects
improves convergence and actually reduces the dependence on the initial
state. Quantum mixing of classical configurations seems to provide bridges
in configuration space for the system to find low energy states. In general,
the quantum results for the ground state energy appear to extrapolate to the
classical results as $t$ is reduced, though the relatively large values of $t
$ to which the Hartree calculations are restricted make this somewhat
difficult to verify in detail. We investigated the region $0<W<2$, $0.3\leq
t<2$ by this method.

\begin{figure}[tbp]
\centerline{\includegraphics[scale=.95,angle=-90]{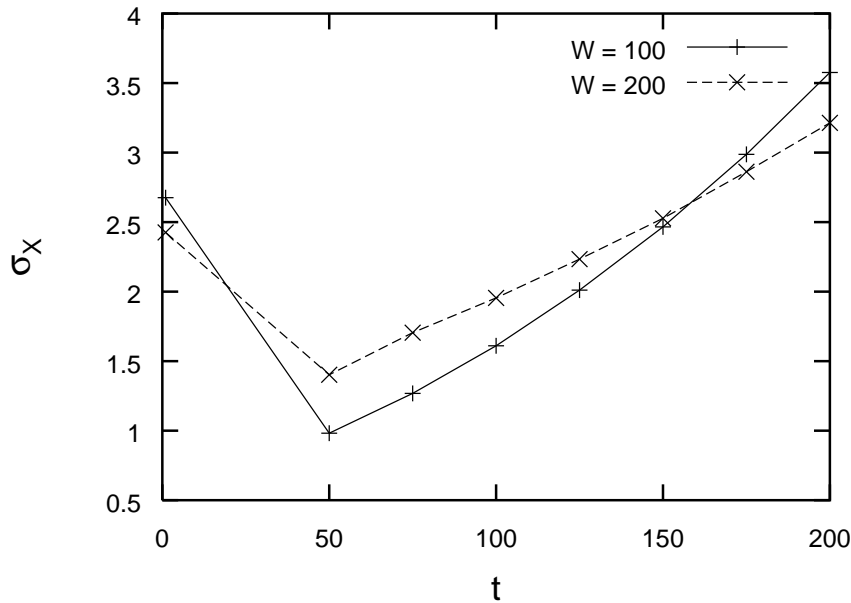}} 
\caption{The width of the CELS distribution as a function of $t$. At small $%
t,$ the width drops quickly, but starting around $t=50$, the width increases
roughly linearly.}
\end{figure}

\begin{figure}[tbp]
\centerline{
\begin{tabular}{cc}
\includegraphics[
  scale=0.65,
  angle=-90]{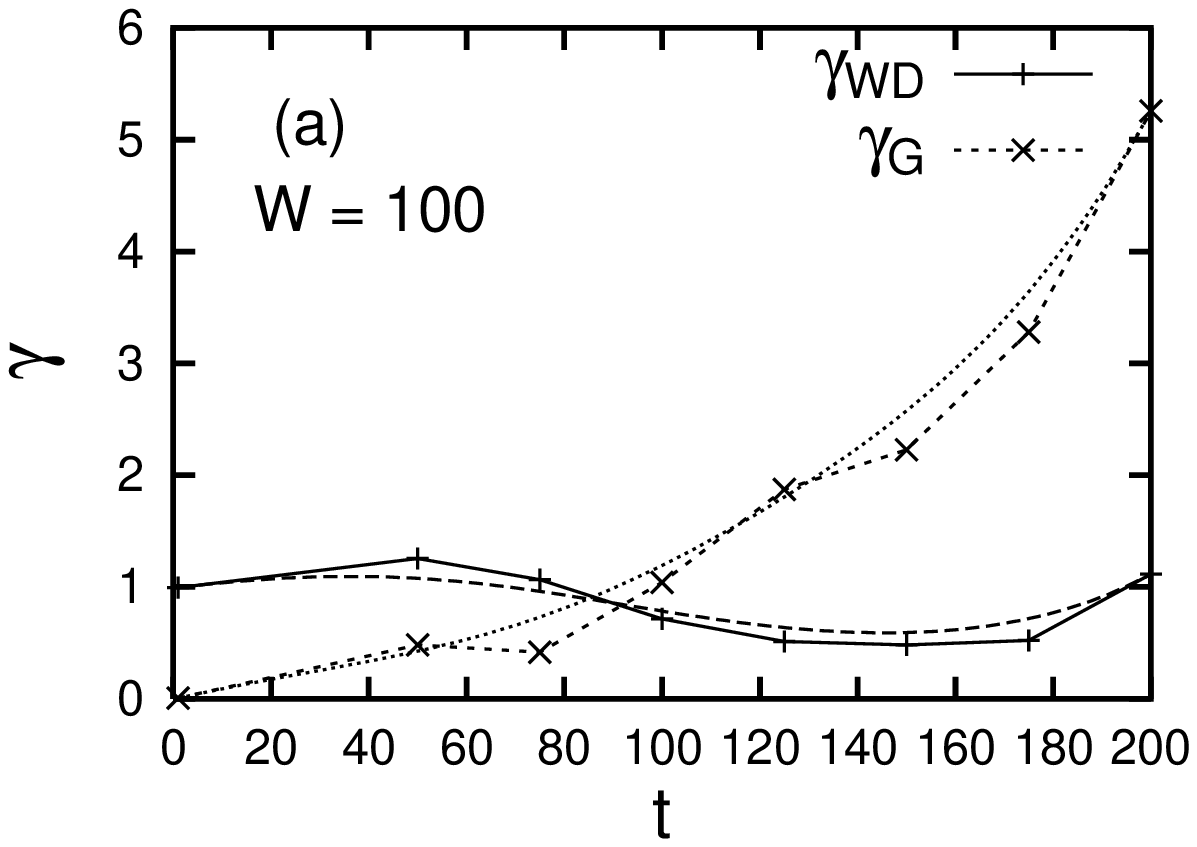}&
\includegraphics[
  scale=0.65,
  angle=-90]{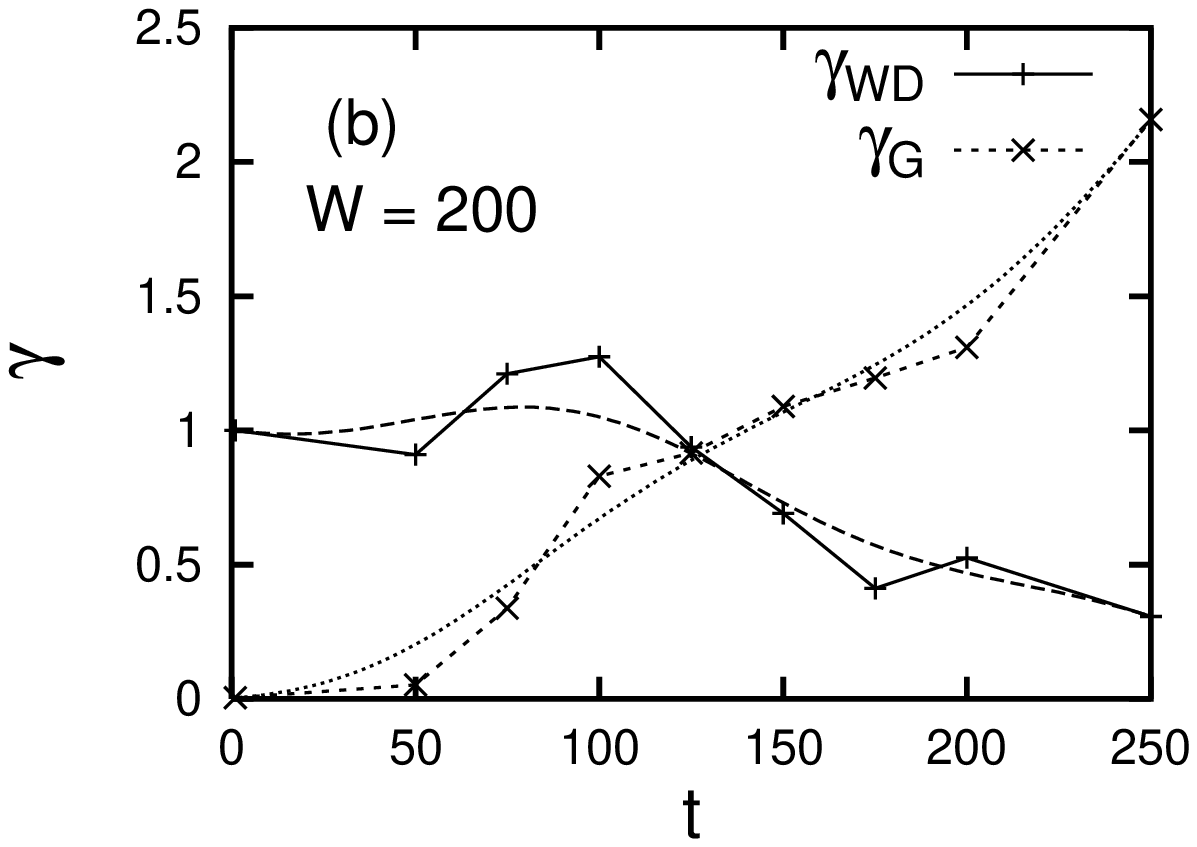}\tabularnewline
\end{tabular}
} 
\caption{Goodness of fit $\protect\gamma $ as a function of $t,$ the hopping
parameter at strong disorder. Small $\protect\gamma $ indicates a good
fit.  $\protect\gamma _{WD}$ and $\protect\gamma _{G}$ refer to
Wigner-Dyson and Gaussian distribution. We see evidence of a crossover
from the Gaussian to Wigner-Dyson as $t$ increases. The +'s and $\times $'s
are numerical data whereas the solid and dasted lines are smoothed data. The
crossover points are (a) $t\approx 85$ for $W=100$ and (b) $t\approx 125$
for $W=200$.}
\end{figure}

We begin by looking at $\sigma _{X}(W,t)$, the rms width of $P(\Delta _{2}).$
\ At $t=0$, (classical case), Fig. 3 has already shown that $\sigma _{X}$ is
proportional to $W$. However, a rather surprising result emerges
immediately, which is a very rapid decline in $\sigma _{X}$ with $t$ even at
small $t$, for fixed $W$. This effect gets weaker as $W$ increases, but it
is very pronounced for all $W<2.$ \ The correlations that are the
consequence of energy level repulsion turn on at small values of $t$: at $%
t=0.3$, the width is considerably less than at $t=0$ and, on further
increasing $t$, $\sigma _{X}$ flattens out and becomes quite small. \ Thus,
from the point of view of the CELS, the system becomes quantum-mechanical at
remarkably small values of the hopping. The decline in $\sigma _{X}$ cannot
continue indefinitely, since ultimately non-universal effects will take
over. This is illustrated in Fig. 6, where we see a break in slope at about $%
t=50,$ where the Wigner-Dyson description breaks down. This is certainly out
of the range of interest for experiments.

\ Unfortunately, the above-mentioned difficulty of obtaining convergence at
small $t$ makes it difficult to describe in detail the leading behavior of $%
\sigma _{X}(W,t)$ at small $t$. However, to the extent that we can
extrapolate the data from finite $t$ to $t=0$, they appear to join smoothly.
Since the methods by which the points are obtained are quite different, this
gives us confidence in the numerical results. (It would be interesting in
future to combine the genetic algorithm with the Hartree approximation and
iteration scheme.)

We now investigate the crossover from Gaussian to Wigner-Dyson CELS as a
function of hopping strength - one might think of increasing $t$ as
continuously strengthening the quantum character of the system. We define
the goodness-of-fit parameters analogously to those defined above:%
\[
\gamma _{WD}=\frac{\sum_{i}\left[ P(\Delta _{2i})-P_{WD}(\Delta _{2i})\right]
^{2}}{\sum_{i}\left[ P_{G}(\Delta _{2i})-P_{WD}(\Delta _{2i})\right] ^{2}} 
\]%
and%
\[
\gamma _{G}=\frac{\sum_{i}\left[ P(\Delta _{2i})-P_{G}(\Delta _{2i})\right]
^{2}}{\sum_{i}\left[ P_{G}(\Delta _{2i})-P_{WD}(\Delta _{2i})\right] ^{2}}. 
\]%
In Fig. 7, we give results of the fits as a function of $t$ for different
values of $W$. Alhassid \textit{et al.} were able to show in their system
that $P(\Delta _{2})$ could be described by a convolution of a Gaussian and
a Wigner-Dyson distribution \cite{alhassid2}. This is also consistent with
our results. We find a crossover from the Gaussian CELS at low $t$ to the
Wigner-Dyson CELS\ at a value of $t=50$. This corresponds roughly to $%
r_{s}=5-10$. This is considerably larger than has been found in previously
studies on somewhat different models \cite{gefen} \cite{alhassid}, but the
model of Ref. \cite{gefen} uses short-range interactions and the model of
Ref. \cite{alhassid} uses random interactions. Furthermore, our criterion
for the crossover uses a more flexible fit for the Gaussian distribution
than for the Wigner-Dyson distribution, possibly pushing the crossover to
higher $r_{s}$.

\begin{figure}[tbp]
\centerline{ 
\begin{tabular}{cc}
{\footnotesize P($\Delta_{2}$)}\begin{tabular}{c}
\includegraphics[scale=0.6]{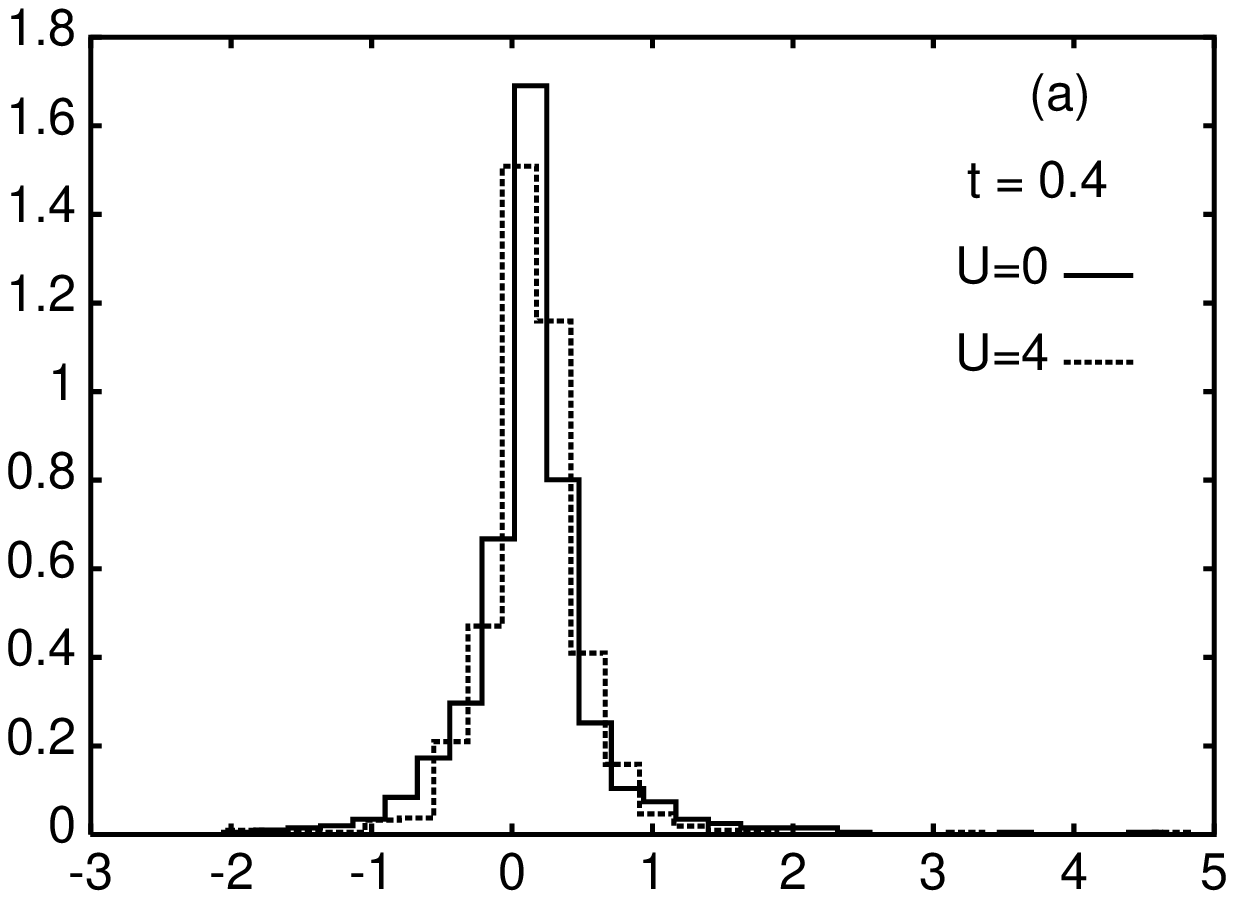}\tabularnewline
{\footnotesize $\Delta_{2}$}\tabularnewline
\end{tabular}&
{\footnotesize P($\Delta_{2}$)}\begin{tabular}{c}
\includegraphics[scale=0.6]{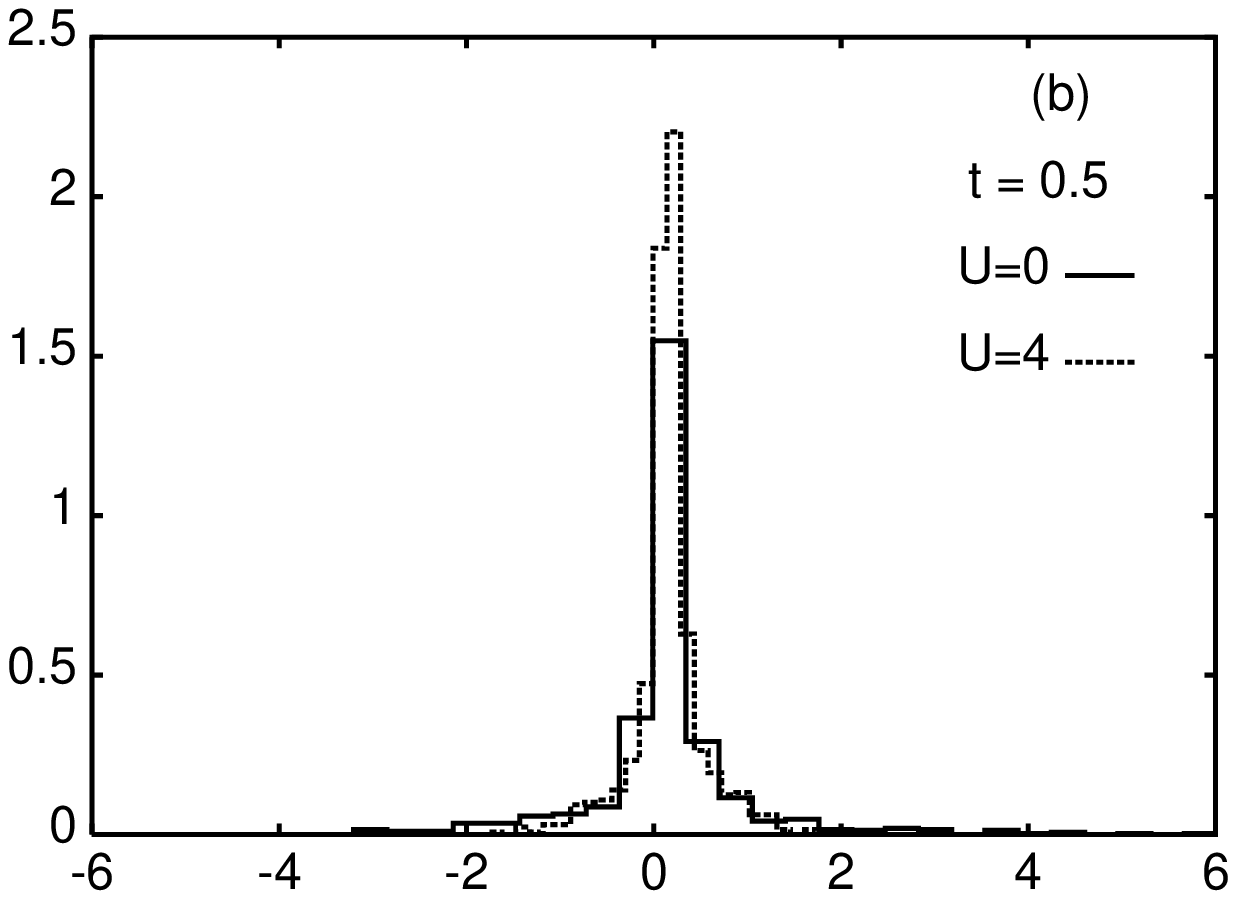}\tabularnewline
{\footnotesize $\Delta_{2}$}\tabularnewline
\end{tabular}\tabularnewline
\end{tabular}
} 
\caption{The effect of on-site interaction on $P(\Delta _{2})$ at $W=2.0$
and (a) $t=0.4$, (b) $t=0.5$. There is little difference in both the form of
the distribution and the width.}
\end{figure}

In the Hartree approximation, antisymmetry of the wavefunction under
particle exchange is not enforced. This means that two particles can be on
the same site. We then need to give a number for the onsite interaction. \
We implicitly chose this as $U=4$ in the calculations so far. However, we
tested the change in the results when $U$ is varied in some test cases. \
The results are shown in Fig. 8. The effect of $U$ is generally quite small,
as one would expect in this range of density.

Finally, the system sizes that we can study are relatively small. This means
that errors due to finite-size effects may be important. We could not make a
systematic study of finite-size scaling. However, we did investigate these
effects by studying a limited parameter set on a $16\times 16$ lattice and
found results very similar those on the $20\times 20$ lattice.

\section{Statistics Plot}\label{sec:phase}

\begin{figure}[tbp]
\centerline{\includegraphics{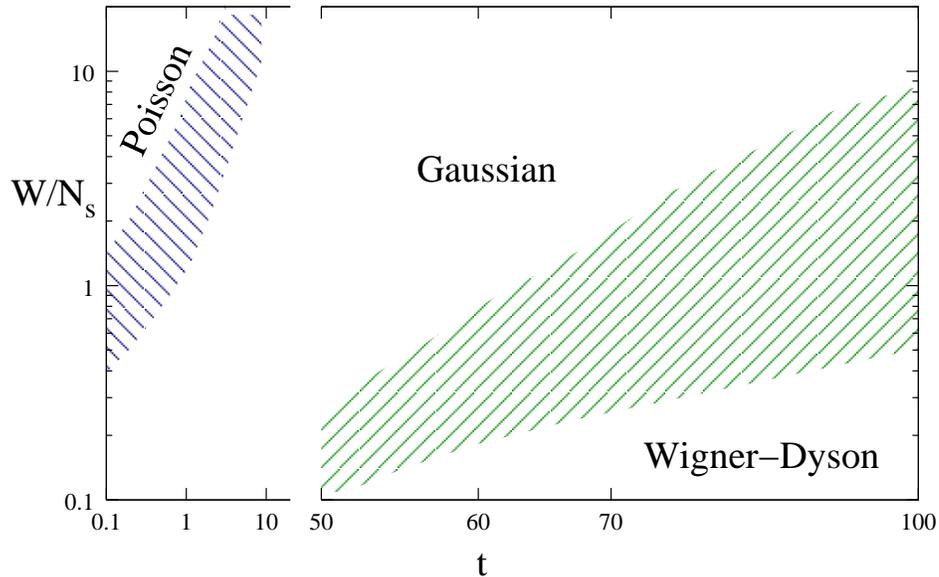}}
\caption{The statistics plot for the peak spacing for our dot model. We
see rather broad crossover regions between the three asymptotic statistics.
 The region between Gaussian and Poisson in particular contains a broad
parameter range where the distribution appears similar to a two-sided
exponential.}
\end{figure}

We may summarize the results of the calculations as follows. In the
classical limit with $t=0$ the effect of classical shells of charges on the
CELS\ is small, even though the rearrangement of charges may be substantial.
\ Thus the width of $P(\Delta _{2})$ is dominated by the disorder even at
quite small $W.$ \ This manifests itself as an increase in $\sigma ,$ the
width of $P(\Delta _{2})$ which is at first linear in $W$ (a finite-size
effect) and then follows a square-root law. There is a crossover from pure
Gaussian to a two-sided exponential-type function for $P(\Delta _{2})$ as $W$
increases. The crossover to a\ true Poisson distribution characteristic of
strongly localized states occurs at much higher $W,$ probably outside the
experimentally accessible range except for the most disordered samples. $%
\sigma $ is, however, very large in the classical case even for relatively
modest values of $W.$ \ When quantum effects are turned on, the width of the
distribution drops precipitously, owing to the usual level-repulsion
effects. If the disorder is small, then the Gaussian $P(\Delta _{2})$ turns
into the Wigner-Dyson form at about $r_{s}\sim 5-10.$ \ For larger disorder,
the two-sided exponential distribution first turns into Gaussian and finally
into Wigner-Dyson. This corresponds to the presence of classical disorder,
interactions, and hopping, respectively. The new effect seen in our work is
the two-sided exponential distribution. This appears because, unlike
previous authors, we analyze the effect of strong disorder. In general, this
effect is to modify the usual Gaussian distribution by first producing long
tails, and then asymmetry in $P(\Delta _{2}).$

In Fig. 9, the results are summarized graphically in the statistics plot, as
determined numerically. Although the numerical results are not definitive,
they suggest that the phase boundaries between the various regions are more
or less straight lines, and that there is always a Gaussian region that
interposes between the Poisson and Wigner-Dyson regimes. Unfortunately,
because of the surprisingly large crossover regions, we cannot completely
resolve the topology of the plot. \ 

\section{Relation to Experiments}

\label{sec:experiments}

\begin{figure}[tbp]
\centerline{\,\includegraphics[scale=.6]{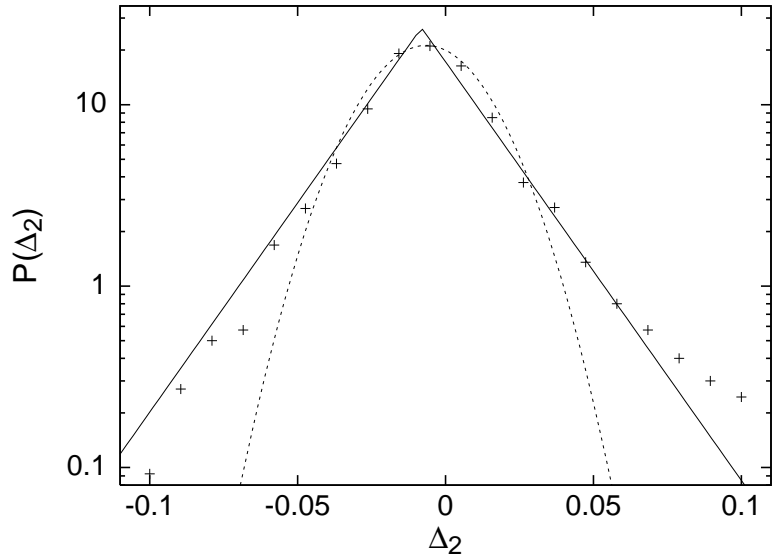}}
\caption{Fitting of the experimental data of Patel et al. (crosses) with
Gaussian (dashed line) and two-sided exponential (solid line) functions. 
The two-sided exponential appears to be somewhat better fit. }
\end{figure}

Several experiments have been performed to measure the CELS. We discuss four
of these.

The most detailed data come from the experiments of Patel \textit{et al}. 
\cite{patel}, who investigated seven GaAs quantum dots, all in the ballistic
regime. Several thousand conductance peaks were examined, in contrast to the
other reports, which contained of the order of one hundred. The mobilities
ranged from ($1.4-6.5)$ $\times 10^{5}$ V-cm$^{2}$/s, and the densities from 
$2-3\times 10^{11}/$cm$^{2}.$ \ $r_{s}\approx 2-3$ for these samples. \ $%
P(\Delta _{2})$ is very symmetric, but with very definite non-Gaussian
tails. These data fit the two-sided exponential quite well, as seen in Fig.
10. The fit is not conclusive, but certainly suggests that there are tails
induced by disorder in these samples. It is of particular interest that
although the samples included have mobilities that vary by a factor of five,
the fit with a single function is still satisfactory. This is probably due
to the fact that the crossover region between Gaussian and Poisson
statistics is almost vertical, as can be seen in Fig. 9.

In the experiment of Sivan \textit{et al.} \cite{sivan}, the system was also
a GaAs quantum dot, coming from a relatively high-mobility\ ($5\times 10^{5}$%
V-cm$^{2}/$ s) sample with density $n_{s}=3.1\times 10^{11}/$ cm$^{2}$. \
Thus the sample was very similar in terms of its disorder and density to
those of Patel \textit{et al.} $P(\Delta _{2})$ is again symmetric, but the
statistics are insufficient to decide on whther the tails are non-Gaussian.

Other Coulomb blockade data come from Ref. \cite{chandrasekhar}, 
as extracted in Ref. \cite{sivan}. This is a different system, consisting of
In$_{2}$O$_{3-x}$ wires that are insulating in the bulk. Unlike the quantum
dots, these systems are believed to be well into the diffusive regime. There
are relatively few accessible quantum states and the disorder is probably
much stronger than in the other sample. There is some evidence of asymmetry
in these data, a possible indication that this system belongs closer to the
Poisson regime on the statistics plot, where the distribution becomes truly
asymmetric.

Finally, we discuss the experiments of Simmel \textit{et al.} \cite{simmel}.
\ These experiments were performed on a Si quantum dot. They are
distinguished from the GaAs dots by a larger $r_{s}.$ \ We would expect the
effects of disorder to be more pronounced. As in all the experiments except
those of Patel \textit{et al}., there are relatively few points, and it is
therefore impossible to judge whether long tails are present. We note from
Fig. 10 that asymmetry is much more likely to show up when $r_{s}$ increases
($t$ decreases in the figure). Given this, the suggestion of asymmetry in
the data may indicate that the system is close to the Poisson crossover.

\section{Conclusion}

\label{sec:conclusion}

We have considered the effect of disorder and interactions on the CELS of
quantum dots with a view towards obtaining a global picture of the CELS. \
We computed the measurable quantity $P(\Delta _{2})$ numerically, averaging
it over many realizations of the disorder. The chief new result is that
strong disorder can modify the Gaussian statistics by producing first
non-Gaussian tails and then asymmetry (skewness) in the distribution,
leading ultimately to the Poisson distribution expected at very stroing
disorder. These is good evidence for the tails and suggestions of the
asymmetry, though relatively poor statistics makes it difficult to say that
these effects have been unambiguously seen. We also find the expected
crossover from Gaussian to Wigner-Dyson statistics as a function of $r_{s}.$
\ Our calculations, which include the long-range Coulomb interaction,
suggest that the crossover occurs at somewhat larger $r_{s}$ than previous
calculations on short-range models have given.

We thank S.N. Coppersmith, M.E. Eriksson, D.D. Long, and N.V. Lien for
useful conversations and acknowledge the support of the National Science
Foundation through Grant No. OISE-0435632, ITR-0325634, and the Army
Research Office through Grant No. W911NF-04-1-0389.

\end{document}